# A review and prospects for Nb$_3$Sn superconductor development

Xingchen Xu

Fermi National Accelerator Laboratory, Batavia, Illinois, US 60510

E-mail: xxu@fnal.gov

**Abstract**

Nb$_3$Sn superconductors have significant applications in constructing high-field (> 10 T) magnets. This article briefly reviews development of Nb$_3$Sn superconductor and proposes prospects for further improvement. It is shown that significant improvement of critical current density ($J_c$) is needed for future accelerator magnets. After a brief review of the development of Nb$_3$Sn superconductors, the factors controlling $J_c$ are summarized and correlated with their microstructure and chemistry. The non-matrix $J_c$ of Nb$_3$Sn conductors is mainly determined by three factors: the fraction of current-carrying Nb$_3$Sn phase in the non-matrix area, the upper critical field $B_{c2}$, and the flux-line pinning capacity. Then prospects to improve the three factors are discussed respectively. An analytic model was developed to show how the ratios of precursors determine the phase fractions after heat treatment, based on which it is predicted that the limit of current-carrying Nb$_3$Sn fraction in subelements is ~65%. Then, since $B_{c2}$ is largely determined by the Nb$_3$Sn stoichiometry, a thermodynamic/kinetic theory was presented to show what essentially determines the Sn content of Nb$_3$Sn conductors. This theory explains the influences of Sn sources and Ti addition on stoichiometry and growth rate of Nb$_3$Sn layers. Next, to improve flux pinning, previous efforts in this community to introduce additional pinning centers (APC) to Nb$_3$Sn wires are reviewed, and an internal oxidation technique is described. Finally, prospects for further improvement of non-matrix $J_c$ of Nb$_3$Sn conductors are discussed, and it is seen that the only opportunity for further significantly improving $J_c$ lies in improving the flux pinning.

**Keywords:** $Nb_3Sn$, $J_c$, area fraction, stoichiometry, flux pinning, APC.

1. Introduction

As the first superconductor that demonstrated the capability to carry high loss-less current at high magnetic fields [1], $Nb_3Sn$ is known as the material that transformed superconductivity from a scientific curiosity to an area of significant practical importance by opening the door to high-field magnet applications [2]. Although the last few decades witnessed the domination of the ductile Nb-Ti in superconducting magnet market, in recent years massive use of $Nb_3Sn$ has expanded dramatically as fields beyond the limit of Nb-Ti are needed. As an example, over 500 tonnes of $Nb_3Sn$ strands were procured by the international thermonuclear experimental reactor (ITER) project from 2008 to 2015, driving a ten-fold increase of worldwide $Nb_3Sn$ production capability [3]. With critical temperature ($T_c$) up to 18.3 K, upper critical field ($B_{c2}$) around 30 T, and high critical current density ($J_c$) at high fields (e.g., whole-wire $J_c$ of 1000 A/mm$^2$ at 4.2 K, 15 T [4]), $Nb_3Sn$ fills the gap between Nb-Ti and the costly and not-yet-mature high temperature superconductors (HTS).

Current and potential application areas of $Nb_3Sn$ superconductors include: dipole and quadrupole magnets for particle accelerators, central solenoids (CS) and toroidal field (TF) coils in tokamak fusion devices (such as ITER [3]), magnetic resonance imaging (MRI) [5], nuclear magnetic resonance (NMR) [6], and laboratory-used high-field magnets including hybrid magnets, and so on. Non-magnet applications, such as superconducting radio frequency (SRF) cavities [7], are also promising.

For all magnet applications, high $J_c$ is an important requirement in order to make magnets compact and economical. Apart from high $J_c$, the critical operation conditions of superconducting magnets put several other requirements on $Nb_3Sn$ superconductors. First, to combat instability, $Nb_3Sn$ conductors are typically produced in the form of round wires, each comprised of a plurality of fine superconducting units ("filaments" or "subelements") embedded in a Cu matrix. Two parameters have been found critical for the electromagnetic stability of superconductors: the subelement size $d_{sub}$, and the thermal conductivity of the Cu matrix, which at low temperatures is sensitive to its purity. The latter is gauged by the residual resistivity ratio (RRR) usually defined as the electrical resistivity at 273 K divided by that at 20 K. Small filament size and high RRR are desired for $Nb_3Sn$ stability [8,9]. Second, to reduce field-ramp-rate-dependent a.c. loss, the superconductors should have small filament size and the filaments should be twisted. Third, reduction of the parasitic magnetic field produced by screening current induced by external magnetic field also requires small $d_{sub}$ [10]. A summary of the requirements on $Nb_3Sn$ conductors for different applications is shown in Table I. Apart from requirements on $J_c$ and $d_{sub}$, all applications require high RRR and good stress/strain tolerance as well.

Table I. The main characteristics of superconducting magnets and their requirements on $Nb_3Sn$ conductors

| Applications in | Targeted Field, T | Field ramp rate, T/s | Requirement on $J_c$ | Requirement on $d_{sub}$ |
|---|---|---|---|---|
| NMR | Up to 24 | Low | $J_e \geq 100$ A/mm² at operational fields | < 100 µm [4] |
| Accelerators | 10 - 16 | ~0.01 | Non-Cu $J_c \geq 1500$ A/mm² at 4.2 K, 16 T* | ≤ 50 µm |
| Fusion | 11 - 13 | 0.1-10 | Non-Cu $J_c \geq 800$ A/mm² at 4.2 K, 12 T** | Well below 10 µm |

\* The example of the planned Future Circular Collider (FCC) [11] is used.
\*\* The example of the ITER project is used [12].

Among these applications, the next biggest potential market for $Nb_3Sn$ superconductors is in accelerator magnets. For example, the planned Future Circular Collider (FCC) requires about eight thousand tons of $Nb_3Sn$ strands [13]. The need to boost particle collision energy requires dipole magnets with higher field for bending particle beams, with the targeted operational field for FCC being 16 T. For a cosine-theta type dipole magnet, the magnetic field *B* generated by a coil is determined by the coil width (*W*) and electric current density in the coil ($J_{coil}$): $B \propto W \bullet J_{coil}$ [14]. With a chosen operational margin, $J_{coil}$ is determined by the $Nb_3Sn$ conductors' $J_c$ and fill factor in the coil. Since the coil size $A_{coil} \propto (W^2+a \bullet W)$, where *a* is the aperture diameter, a large *W* is undesirable because it causes sharp increase of amount and cost of conductors and other materials, and challenges in quench protection of magnets. To restrain *W*, high-$J_c$ conductors are required. A minimum $J_c$ requirement has been put forward for the FCC conductors [11]. A comparison between the FCC requirement and the present rod-restack-process (RRP) wires [16], which represent the state-of-the-art level of $Nb_3Sn$ conductors, is shown in Figure 1. Although the highest 15 T $J_c$s of top-performing RRP conductors can reach 1600-1700 A/mm$^2$ and are only 15-20% below the FCC specification, due to the inevitable variations among billets in series production [15], at least 40-45% improvement is still required to fulfill the FCC requirement.

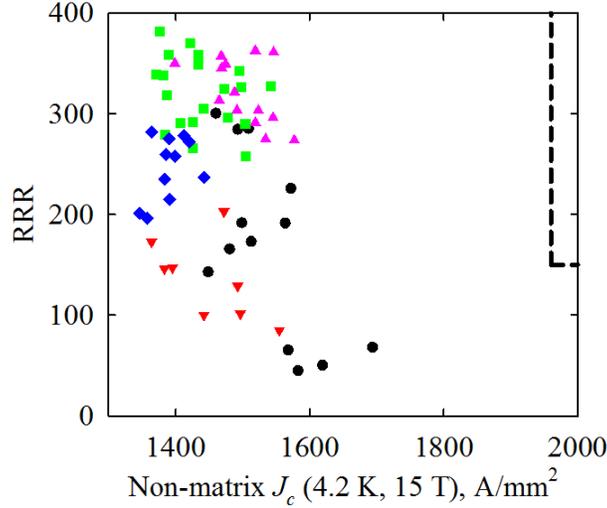

Figure 1. RRR vs 15 T $J_c$ for RRP wires produced for the Large Hadron Collider (LHC) Luminosity Upgrade [16]. Different symbols represent different designs (see [16] for details). All wires are Ti doped, with $d_{sub}$ of 50-55 μm, and heat treated at 650-665 °C. The horizontal and vertical dashed lines denote the minimum RRR and $J_c$ levels required by the FCC specification [11].

On the other hand, the state-of-the-art $J_c$s of Nb$_3$Sn conductors have plateaued since the early 2000s [17] (the improvement of Nb$_3$Sn record $J_c$s for long-length wires with time is shown in Figure 2 [18]), in spite of the significant efforts that have been made to improve them. This article will discuss prospects for further improvement of Nb$_3$Sn superconductors.

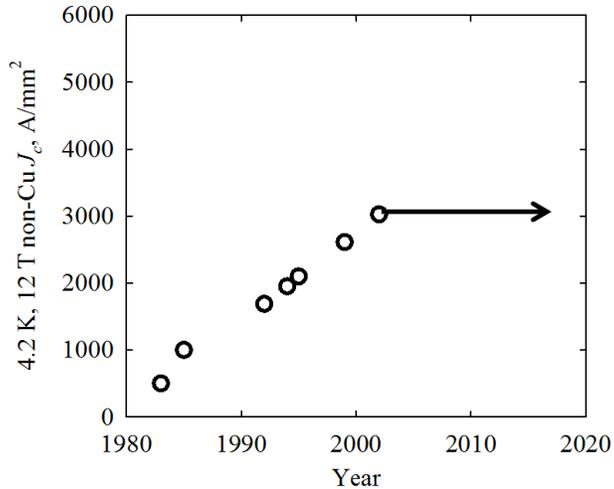

Figure 2. Improvement of the record 12 T $J_c$ of Nb$_3$Sn conductors with time [18].

## 2. Review of Nb$_3$Sn conductor development

Nb$_3$Sn was found to be superconducting in 1954 [19]. Soon after the publication of Kunzler's exciting results in 1961 [1], 7-15 T Nb$_3$Sn magnets were made commercially available [20]. These early magnets were wound from Nb$_3$Sn tapes with thin Nb$_3$Sn layers deposited or formed on metallic substrates [21,22], which were flexible enough for winding into coils. The application of Nb$_3$Sn tapes, however, was discouraged by the problem of electromagnetic instabilities, particularly flux jumping at low fields. It was then realized in the late 1960s that such instabilities could be suppressed by sub-dividing the superconductors [23]. This fostered the development of multi-filamentary Nb$_3$Sn wires with fine superconducting units embedded in a metallic matrix with high thermal conductivity.

Thanks to the discovery that the participation of Cu in the Nb-Sn reaction facilitated the formation of Nb$_3$Sn at relatively low temperatures [24], the first multi-filamentary Nb$_3$Sn wires were produced using the so-called "bronze-process" method around 1970 [25,26]. In a typical process using this method, a certain number of Nb rods are inserted into the axially drilled holes in a bronze billet, and the composite is wrapped with a tantalum foil and inserted into a Cu can, and the whole composite is drawn down to the final size. Several intermediate anneals between drawing courses are required to relieve work hardening of the bronze. The final-size, green-state wires are wound to coils, and finally, heat treatments are applied during which bronze reacts with Nb to form $Nb_3Sn$ phase. The Ta layer protects the outer Cu matrix from contamination by the bronze during heat treatment. In this type of wire, bronze with higher Sn content leads to higher $Nb_3Sn$ performance [27], but the Sn content in bronze is limited by its processibility.

To avoid the frequent annealing during wire drawing, an "internal-tin" approach was developed in 1974 [28]: numerous Nb rods were inserted into axial holes drilled in a Cu billet, and a few Sn rods were distributed among Nb filaments to supply Sn, with scanning electron microscopy (SEM) images shown in Figure 3 (a). During heat treatment the Cu and Sn first mix to form Cu-Sn alloys, which then react with Nb to form the $Nb_3Sn$ phase. Compared with bronze-process wires, the internal-tin type can have Sn/Cu/Nb ratios varied over a large range.

In both bronze-process and single-barrier internal-tin strands, each Nb rod transforms to a $Nb_3Sn$ filament after heat treatment; due to the large Cu/Nb ratio (which causes large spacing among the Nb rods), the filaments are generally separated by low-Sn bronze after reaction, leading to small filament size (several μm) and thus low a.c. loss. The price for the high Cu fraction that is necessary for dividing the filaments, is a relatively low $J_c$. Such strands find significant applications in fusion devices, laboratory magnets and NMR magnets.

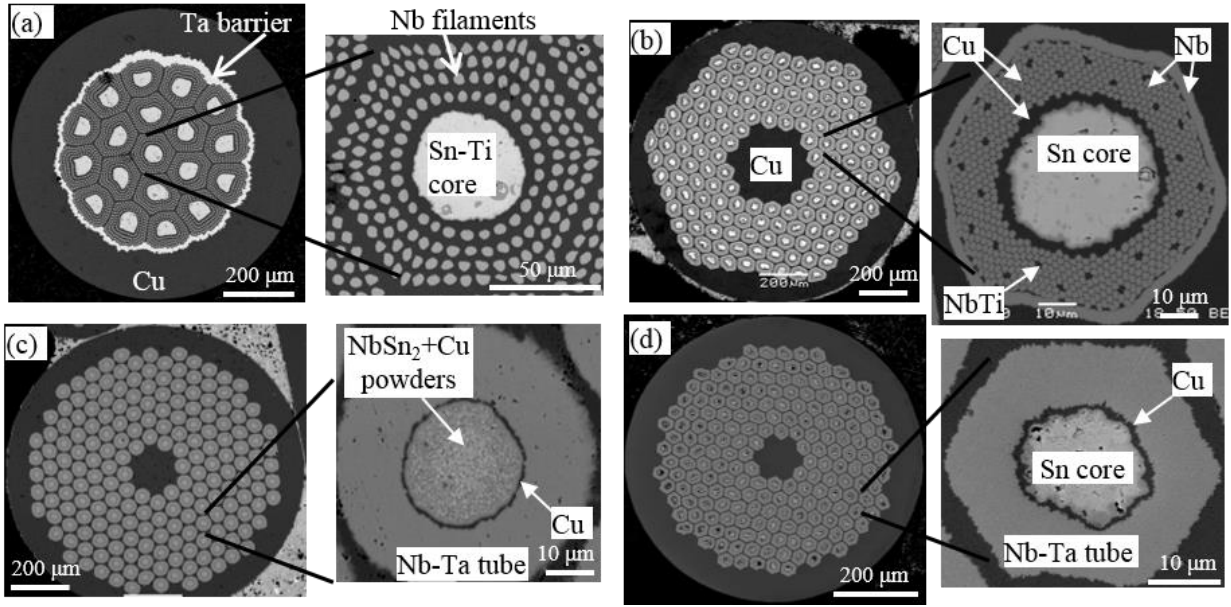

Figure 3. SEM images of cross sections of (a) single-barrier internal-tin, (b) RRP, (c) PIT, and (d) TT Nb$_3$Sn strands.

The 1970s and 1980s witnessed the blooming of researches in Nb$_3$Sn superconductor; a large number of studies were made in this period to explore what controlled the superconducting properties of Nb$_3$Sn. The Nb-Sn binary phase diagram was determined [29], which revealed the presence of Nb$_6$Sn$_5$ and NbSn$_2$ phases and the composition range of Nb$_3$Sn phase (~17-25.5 at.% Sn). At this time researchers also realized that the $T_c$ and $B_{c2}$ depend strongly on the stoichiometry of Nb$_3$Sn – deficiency in Sn was shown to lead to seriously degraded $T_c$ and $B_{c2}$ [30,31]. The $B_{c2}$ was also correlated with the normal state resistivity [32], stimulating researchers to study improvements to $B_{c2}$ by way of doping. Numerous studies were made regarding the influence of various additions on the superconducting properties of Nb$_3$Sn (for example, [33-36]). These led to the selection of tantalum and titanium as dopants [36], which are routinely used for all present-day Nb$_3$Sn strands. In addition to the studies about Nb$_3$Sn chemistry and superconducting properties, researchers also made efforts to optimize microstructures in order to improve $J_c$. It was found that grain boundaries are the primary flux pinning centers for $Nb_3Sn$ and that reduction in grain size is a very effective approach to improving pinning force and $J_c$ [37,38].

Also during this time some variants of internal-tin strands were developed. One of them is the modified jelly roll (MJR) approach [39], in which a subelement is produced by rolling laminated Cu foil and expanded Nb mesh around a central Cu-clad Sn rod. Later distributed barriers were introduced to MJR wires so that each subelement had its own barrier against Sn leakage into the outer Cu matrix [40]. This turned out to be a very important change for improving $J_c$ of $Nb_3Sn$ wires, because it allowed for significant decrease of the Cu content inside each barrier (note that the Cu inside the barrier participates in the Nb-Sn reaction), in which case all the Nb filaments in each subelement merge into a single $Nb_3Sn$ annulus. On the other hand, a significant reduction of Cu content is not an option for a single-barrier wire, because merging of all the filaments into a single current-carrying unit in each wire would lead to excessively large $d_{sub}$. Decreasing Cu content inside the barrier turned out to be a very effective way to improve non-matrix $J_c$ of internal-tin strands. Figure 4 [41] shows the variation of non-matrix $J_c$ with the Sn/Cu ratio inside the barrier for internal-tin strands. It should be noted that the Nb/Cu ratio increases accordingly with Sn/Cu ratio so that Sn is consumed to form $Nb_3Sn$. It can be seen that with the increase of Sn (and also Nb) fractions, the non-matrix $J_c$ increases by a factor of 4. It will be shown later that this is not only due to increase of $Nb_3Sn$ fraction as Cu fraction decreases, but also due to improvement of $Nb_3Sn$ phase quality, such as reduction of Sn content gradient and columnar grain morphology. In fact, the improvement of $J_c$ from the 1980s to the early 2000s shown in Figure 2 was mainly driven by reducing Cu content in subelements of internal-tin wires, although other factors (e.g., optimizing heat treatments and doping) also contributed.

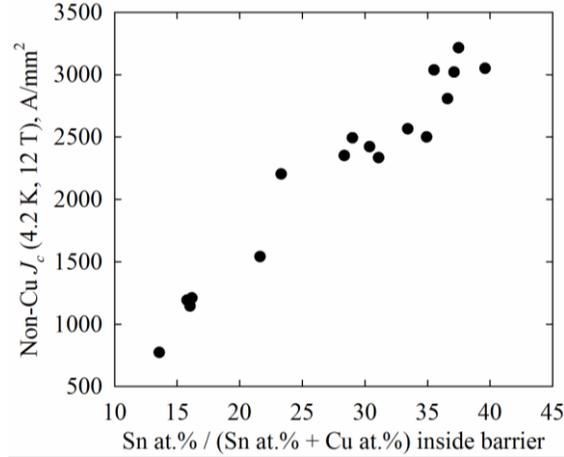

Figure 4. The dependence of non-matrix $J_c$ on Sn/Cu ratio in internal-tin strands [41].

In the early 2000s the MJR approach was replaced by the "rod-restack-process" (RRP) method to facilitate large-scale production [17]. RRP is a distributed-barrier type, and in each subelement a Sn rod is inserted into the central hole of a Cu layer that is surrounded by a stack of Cu-clad Nb rods inside a Nb barrier (Figure 3 b). In the RRP method, the Cu/Nb area ratio (named "local area ratio", LAR) of the Cu-clad Nb rods can be flexibly adjusted. During reaction the Nb filaments transform to a single Nb$_3$Sn annulus in each subelement. In the early 2000s non-Cu $J_c$ of ~3000 A/mm$^2$ was achieved at 4.2 K, 12 T in RRP wires by Oxford Superconducting Technology, with the best values about 3300 A/mm$^2$ [17].

In parallel with the internal-tin type and its variants, other Nb$_3$Sn wire types were also developed. Although the first powder-in-tube (PIT) Nb$_3$Sn wire dated back to Kunzler's work [1], the first multi-filamentary PIT wire with Cu addition to assist Nb-Sn reaction at low temperatures was not developed until 1977 by the Netherlands Energy Research Foundation (ECN) [42]. In such wires a blending of $NbSn_2$ and small amounts of Cu powders is filled into a thin Cu tube, which is inserted into a Nb (or Nb-7.5wt.%Ta) tube to compose a subelement, images shown in Figure 3 (c). In such wires unreacted Nb in each filament serves as the barrier. Later the technology was transferred to ShapeMetal Innovation (SMI), which, by optimizing heat treatment and adding ternary dopants, increased the non-Cu $J_c$ (4.2 K, 12 T) to ~2500 A/mm$^2$ [43], with a record value of 2700 A/mm$^2$ also reported [44]. Such PIT wires are produced by Bruker-EAS at present.

Also in the 1970s another $Nb_3Sn$ wire manufacturing technology emerged, which produces subelements by inserting Cu-clad Sn rods into Nb (or Nb-7.5wt.%Ta) tubes [45,46]. Clearly, this manufacturing process is much simpler than any other counterparts. Over two decades later this approach was picked up by Supercon Inc. [47] and Hyper Tech Research (HTR) Inc. [48], which named this type of strand "internal-tin tube" (ITT) and "tube type" (TT), respectively. Images of such a strand manufactured by HTR is shown in Figure 3 (d). With optimized Cu/Nb/Sn ratios and heat treatments, the highest non-Cu $J_c$ (4.2 K, 12 T) achieved in this type of strand is nearly 2500 A/mm$^2$ by HTR [49].

Compared with bronze-process and single-barrier internal-tin wires, $Nb_3Sn$ wires featured by the distributed-barrier design (including RRP, PIT, and TT) have high $J_c$s due to the much lower Cu fraction inside each subelement, and they are mainly used for accelerator magnets and NMR magnets. Among these three types, PIT and TT are easier to obtain smaller subelement size thanks to their simpler subelement structures. SMI successfully manufactured PIT strands with 20-25 μm subelement size [50], while HTR has successfully produced TT strands with $d_{sub}$ of 12-16 μm and non-matrix $J_c$ (4.2 K, 12 T) nearly 2000 A/mm² [51].

## 3. Factors controlling the Nb₃Sn conductor $J_c$

There are three types of critical current density $J_c$ depending on the area by which the critical current $I_c$ is divided: (1) the engineering $J_c$ (or $J_e$) of a wire is defined as $I_c$ divided by its whole cross-sectional area, (2) the non-matrix $J_c$ is $I_c$ divided by the total Nb₃Sn subelement area (i.e., the whole cross-sectional area minus the Cu matrix area), and (3) the layer $J_c$ is $I_c$ divided by the current-carrying Nb₃Sn layer area. The layer $J_c$ reflects the quality of the Nb₃Sn layer, while the non-matrix $J_c$ equals to the product of the Nb₃Sn layer $J_c$ and the current-carrying Nb₃Sn area fraction in each subelement.

The Nb₃Sn layer $J_c$ at a certain field $B$ is determined by the bulk pinning force density $F_p$ (N/m³) that the pinning centers can provide at this field to balance the Lorentz force on the flux lines: $J_c = F_p/B$. The $F_p$ is a function of $B$, and the $F_p(B)$ for Nb₃Sn has been shown to have the following form at high fields [52-54]:

$$F_p(B) = 1.5625\sqrt{5} F_{p,max}(B/B_{c2})^{1/2}(1 - B/B_{c2})^2 \tag{1}$$

where $F_{p,max}$ is the maximum bulk pinning force density, and $B_{c2}$ (upper critical field) is the field at which $J_c$ vanishes. According to this equation, a Nb₃Sn layer $J_c(B)$ curve is determined by only two parameters − $F_{p,max}$ and $B_{c2}$; increase in $F_{p,max}$ or $B_{c2}$ can both lead to improvement of $F_p$ and $J_c$ in the high-field range ($\geq 0.3 B_{c2}$) where Nb₃Sn superconductors are used. From equation (1) it is clear that $F_p(B)$ curves peak at $0.2B_{c2}$, i.e., $F_p(0.2B_{c2}) = F_{p,max}$. Fitting measured $J_c(B)$ data to equation (1) yields values of $F_{p,max}$ and $B_{c2}$.

It follows that the non-matrix $J_c$ of Nb$_3$Sn superconductors is mainly determined by three parameters: the current-carrying Nb$_3$Sn fraction, $B_{c2}$, and $F_{p,max}$ (assuming the properties of a strand are uniform along its length; otherwise, the $J_c$ of a long strand is limited by that of its weakest segment). These three parameters are closely related to microstructure and microchemistry of conductors. For example, $B_{c2}$ depends on Nb$_3$Sn phase stoichiometry [53], doping [36], and strain state [54], and $F_{p,max}$ mainly depends on volumetric density of flux pinning centers (mainly grain boundaries for Nb$_3$Sn) [54]. Microstructure and microchemistry of strands depend on starting chemistry/structure of the precursors as well as processing and heat treatment. For example, higher heat treatment temperatures lead to higher Sn contents in Nb$_3$Sn which benefit $B_{c2}$, but result in larger Nb$_3$Sn grain sizes which are detrimental to $F_{p,max}$ [55].

An important method to diagnose a low-$J_c$ superconductor is to analyze the above three parameters – this allows for identifying the causes (or limiting factors) for the low $J_c$, which is the basis for improvement. An example is given here. Among the three types of high-$J_c$ Nb$_3$Sn wires (RRP, PIT, and TT) discussed in Section 2, only RRP strands achieve a non-matrix $J_c$ of 3000 A/mm$^2$ at 4.2 K, 12 T, while those of the other two types are typically ~2500 A/mm$^2$ or below. To seek further improvement, a question to ask is, what causes their inferior non-matrix $J_c$s? An explanation was that stoichiometry was relatively uniform across the Nb$_3$Sn layer in a RRP subelement thanks to the Cu networks among Nb filaments that facilitate Sn transportation, while the Sn content gradient across the Nb$_3$Sn layer in a PIT or TT filament was larger.

To find out the true origin, the above three parameters (Nb$_3$Sn fraction, $B_{c2}$, and pinning) of the three types of strands are analyzed below. Figure 5 shows $B_{c2}$ (4.2 K) values obtained from Kramer plots of transport $J_c(B)$ data for RRP, PIT, and TT strands. Clearly $B_{c2}$ increases with reaction temperature, and Ti doped strands have higher $B_{c2}$s than Ta doped strands, although at very high reaction temperatures (e.g., ≥750 °C) they both saturate at a similar level (27-28 T) [56], which is presumed to be close to the limit of $Nb_3Sn$. From Figure 5 it is clear that PIT and TT strands (all with Ta doping) reacted at 615-625 °C have $B_{c2}$s (4.2 K) of 25-26 T, even higher than those of RRP strands. A closer examination of the microchemistry of RRP and TT strands shows that the stoichiometry of the former is not superior to the latter [57].

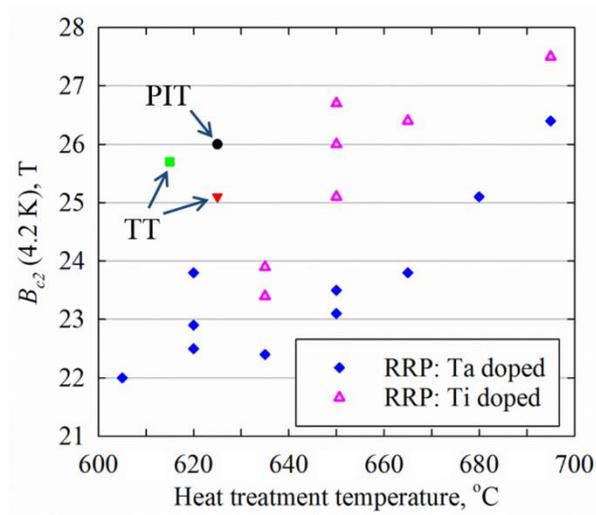

Figure 5. $B_{c2}$ values obtained from Kramer plots of transport $J_c(B)$ data (after self-corrections) for RRP [56], PIT [44], and TT [58] strands. Except an RRP strand with Ti doping (shown in open symbols), all other strands use Nb-7.5wt.%Ta (shown in solid symbols).

Analysis of their microstructures shows that they have similar grain sizes (110-130 nm when reacted at 625-650 °C) and aspect ratios [55,57], which should lead to similar pinning capacities. Hence, it is determined that $B_{c2}$ or layer $F_{p,max}$ should not be the cause of lower 12 T non-matrix $J_c$s for PIT and TT strands relative to RRP. Thus, the $Nb_3Sn$ area fractions may be responsible. Figure 6 shows SEM images of typical TT and RRP subelements after being fully reacted.

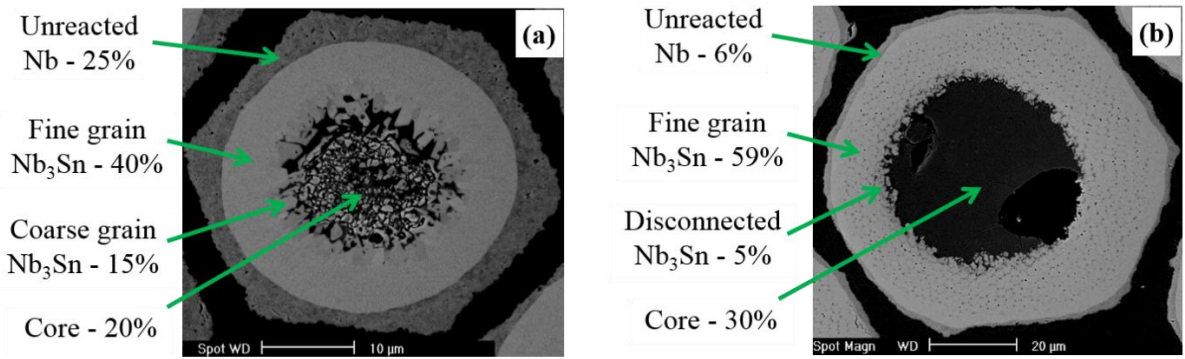

Figure 6. SEM images of fully-reacted (a) TT and (b) RRP subelements with the fractions of components.

As can be seen, a TT subelement is comprised of four components: the remaining Nb area (as a barrier preventing Sn poisoning the outer Cu matrix) typically taking 25-30 % of the subelement, the fine-grain $Nb_3Sn$ area taking 40-45 % of the subelement, the coarse-grain $Nb_3Sn$ area which has large grain size (typically > 1 μm) and takes ~15 %, and a core region (as composed of low-Sn Cu-Sn, Kirkendall voids, and disconnected $Nb_3Sn$ chunks or particles) taking ~20 % of the subelement area. The coarse-grain $Nb_3Sn$ and the $Nb_3Sn$ particles in the core are formed by the dissociation of the intermediate high-Sn Nb-Sn phases, and do not carry supercurrent [59], while the fine-grain $Nb_3Sn$, which is formed by diffusion reaction between Sn and Nb, is the only supercurrent-carrying $Nb_3Sn$ phase in a subelement. Analysis of PIT strands shows similar area fraction components [54]. On the other hand, the fine-grain $Nb_3Sn$ fraction in RRP subelements (Figure 6 b) is much higher (~60%), thanks to the much lower unreacted Nb fraction (6-8 %) and disconnected (or coarse-grain) $Nb_3Sn$ phase (~ 5%) near the core.

Based on the fine-grain $Nb_3Sn$ fractions in subelements, it can be calculated that the 4.2 K, 12 T layer $J_c$ of RRP strands is ~5000 A/mm$^2$, while those of PIT and TT are 5500-6000

A/mm$^2$. Therefore, the origin for the lower non-matrix $J_c$s in PIT and TT strands is not lower quality, but lower quantity, of Nb$_3$Sn phase in subelements; thus, efforts should be placed on improving their fine-grain Nb$_3$Sn fractions.

The next three sections will discuss what determines the three factors (fine-grain Nb$_3$Sn fraction, $B_{c2}$, and $F_{p,max}$) and prospects to improve them, respectively.

## 4. Prospects to improve the Nb$_3$Sn fractions

Based on observations of the phase transformations in TT strands during heat treatments [59], the reaction between Cu-Sn and Nb can be summarized as follows: as η phase (Cu$_6$Sn$_5$) or high-Sn liquid Cu-Sn reacts with Nb, it transforms to ε phase (Cu$_3$Sn) while releasing the extra Sn to form Cu-Nb-Sn ternary phase, which transforms to NbSn$_2$, and then Nb$_6$Sn$_5$ [59,61]; Nb$_6$Sn$_5$ itself transforms to coarse-grain Nb$_3$Sn, releasing the extra Sn for fine-grain Nb$_3$Sn layer growth. After Nb$_6$Sn$_5$ is depleted, the ε phase (or γ phase at high temperatures) then supplies Sn for further fine-grain Nb$_3$Sn layer growth until the chemical potential of Sn in the Cu-Sn source is not higher than that in Nb$_3$Sn. As ε phase or Cu-Sn phase with lower Sn content (e.g., bronze) reacts with Nb, only fine-grain Nb$_3$Sn forms due to the absence of intermediate phases. This is why no coarse-grain Nb$_3$Sn forms in bronze-process or single-barrier internal-tin strands, in which the Cu/Sn ratio is high.

An analytical model was developed to find coarse-grain and fine-grain Nb$_3$Sn amounts for fully-reacted TT strands based on starting amounts of precursors (particularly Sn and Cu, because Nb is in excess so that remaining Nb serves as barrier). The derivation details of the model, which are based on mass conservation, can be found in paper [60]. Expressions for the area inside the barrier ($A_{in}$, not including the barrier area), the fine-grain area ($A_{FG}$), the coarse-grain areas ($A_{CG}$), and the core area ($A_{core}$), in terms of the areas of Sn and Cu in the green-state subelement (noted as $A_{Sn}$ and $A_{Cu}$, respectively), are given in Equations 2-5. Note that these equations are transformed from the equations in [60].

$$A_{in} = (1 + 3.3\, V_m^{Nb}/V_m^{Sn}) \cdot A_{Sn} + [1 + 0.2\, \chi/(1-\chi) \cdot V_m^{Nb}/V_m^{Cu}] \cdot A_{Cu} \quad (2)$$

$$A_{FG} = 0.6\, V_m^{FG}/V_m^{Sn} \cdot A_{Sn} + 0.4\, \chi/(1-\chi) \cdot V_m^{FG}/V_m^{Cu} \cdot A_{Cu} \quad (3)$$

$$A_{CG} = 0.4\, V_m^{CG}/V_m^{Sn} \cdot A_{Sn} - 0.4\, \chi/(1-\chi) \cdot V_m^{CG}/V_m^{Cu} \cdot A_{Cu} \quad (4)$$

$$A_{core} = \left[1 + \frac{3.3 V_m^{Nb} - 0.6 V_m^{FG} - 0.4 V_m^{CG}}{V_m^{Sn}}\right] \cdot A_{Sn} + \left[1 + \frac{\chi}{5(1-\chi)} \frac{V_m^{Nb} + 2 V_m^{CG} - 2 V_m^{FG}}{V_m^{Cu}}\right] \cdot A_{Cu} \quad (5)$$

In equations 2-5, $V_m^{Nb}$, $V_m^{Sn}$, $V_m^{Cu}$, $V_m^{FG}$, $V_m^{CG}$ represent the molar volumes of Nb, Sn, Cu, fine-grain $Nb_3Sn$ and coarse-grain $Nb_3Sn$, respectively, and $\chi$ represents the Sn content in the Cu-Sn phase as $Nb_6Sn_5$ serves as Sn source, which is around 0.25. It should be noted that coarse-grain $Nb_3Sn$ is typically stoichiometric, while fine-grain $Nb_3Sn$ has a Sn content of 22-24 at.%, so $V_m^{FG}$ and $V_m^{CG}$ are not necessarily the same. The molar volumes of different materials can be found in [60]. From the Equation (4), it is clear that $A_{CG}$ decreases with $A_{Cu}/A_{Sn}$, reaching zero at an $A_{Cu}/A_{Sn}$ that is equivalent to a Cu-Sn composition of Cu-$\chi$ Sn. A similar conclusion applies to PIT strands, whose cores contain three elements – Nb, Sn, Cu: as long as the composition of the core makes the formation of $Nb_6Sn_5$ thermodynamically unstable, the formation of coarse grains can be avoided, as shown in the isothermal section of the ternary Cu-Nb-Sn phase diagram in Figure 7.

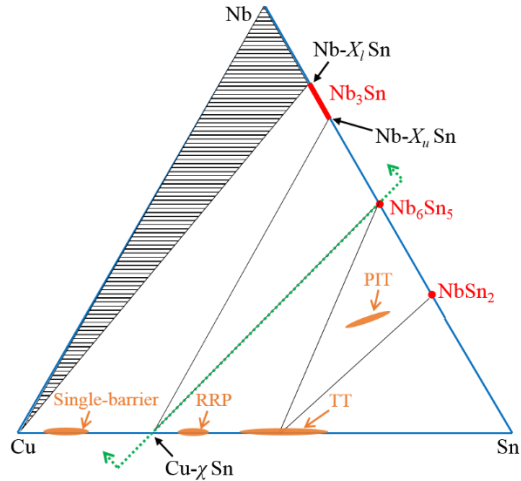

Figure 7. Schematic isothermal section of the ternary Cu-Nb-Sn phase diagram [62]. The green dashed line marks the boundary of the core composition in a PIT wire to the left of which no coarse grains would form. The approximate compositions of Sn sources in PIT, TT, RRP, and single-barrier (including bronze-process and internal-tin) types of $Nb_3Sn$ wires are marked in the figure. Note that details of Cu-Sn phases and some phase equilibrium between Nb-Sn and Cu-Sn phases are not shown here as they are irrelevant for discussions.

To test the above analytic model, TT wires with a large range of Cu/Sn ratios were fabricated and fully reacted, and the fine-grain and coarse-grain $Nb_3Sn$ area fractions were calculated, and it was found that the experimental results agreed well with the predictions of equations 2-5 [60]. As examples, three wires with different Cu/Sn ratios are shown in Figure 8. It is clear that as Cu/Sn ratio increases, the amounts of Cu-Nb-Sn phase and coarse-grain $Nb_3Sn$ decrease while fine-grain $Nb_3Sn$ fraction increases, with details given in [60].

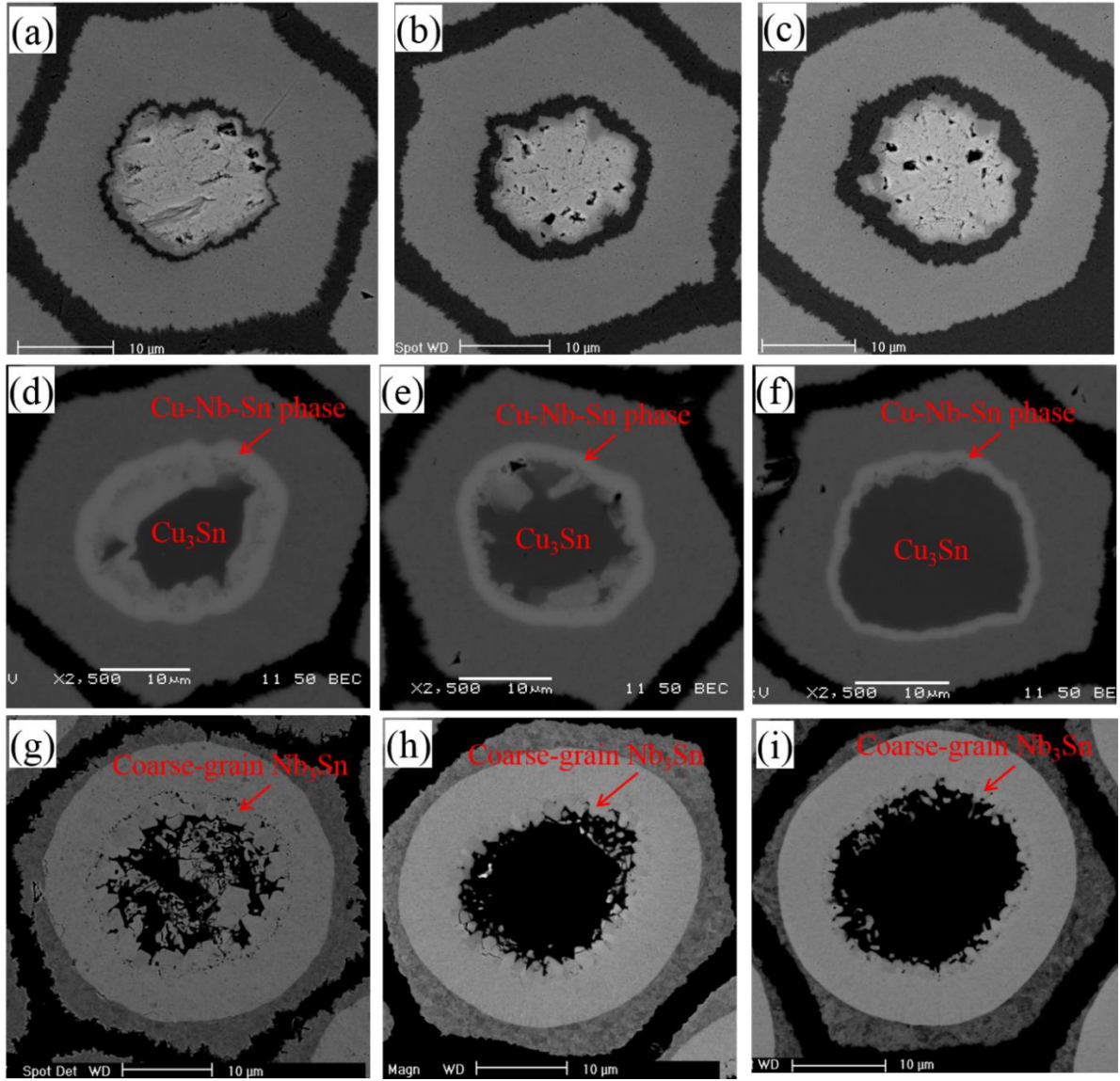

Figure 8. SEM images of the TT wires with low, medium, and high Cu/Sn ratios: (a)-(c) in the green state, (d)-(f) after reaction at 470 °C for 24 hours, (g)-(i) after reaction at 625 °C for 300 hours, respectively.

Figure 9 shows the fractions of fine-grain $Nb_3Sn$, coarse-grain $Nb_3Sn$, and core area in $A_{in}$ as a function of Cu/Sn ratio, calculated from equations 2-5. It can be seen that although the increase of Cu/Sn ratio leads to a marked increase in the core area, the fine-grain $Nb_3Sn$ fraction still increases due to the decrease of coarse-grain amount. According to Figure 9, the fine-grain Nb$_3$Sn fraction is at its maximum when the precursor ratios reach the point where the coarse-grain Nb$_3$Sn amount drops to zero (i.e., no high-Sn intermediate phases form). A larger Cu/Sn ratio relative to this point simply dilutes the fraction of Nb$_3$Sn area due to the excess Cu, as in bronze-process and single-barrier internal-tin wires. This model also applies to RRP wires as phase transformations do not depend on subelement structures. It can be seen that the predicted area fractions in Figure 9 agree reasonably well with those measured ones in Figure 6.

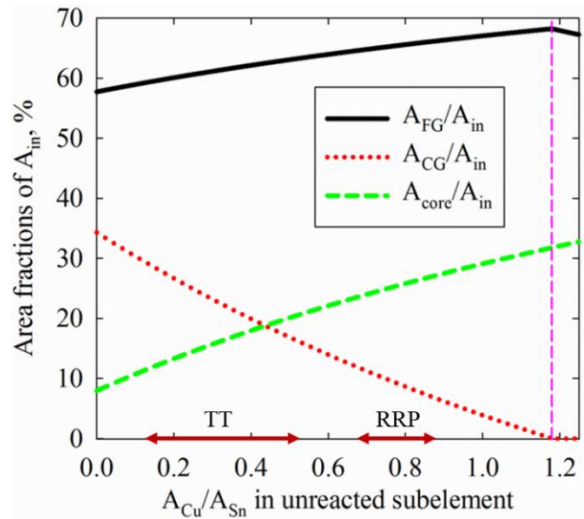

Figure 9. Fractions of fine-grain Nb$_3$Sn, coarse-grain Nb$_3$Sn (including coarse-grain Nb$_3$Sn layer and disconnected Nb$_3$Sn chunks/particles in the core), and the core relative to $A_{in}$ as functions of Cu/Sn ratio, calculated from equations 2-5. The approximate Cu/Sn ratio ranges of TT and RRP wires are shown in the figure as well.

Assuming that the barrier fraction in Nb$_3$Sn subelements can be as small as 5% to preserve RRR (while this value for present RRP is 6-8%), then according to Figure 9, it can be estimated that the limit of the fine-grain Nb$_3$Sn fraction in a subelement is about 68% × 95% ≈ 65 %.

As shown in Figure 6, the fine-grain Nb$_3$Sn fraction in RRP subelements is around 60%, so the room for further improving the Nb$_3$Sn fraction in RRP strands is marginal. New intermediate heat treatments are proposed by Sanabria et al. to reduce the formation of Cu-Nb-Sn ternary phase and thus improve the fine-grain Nb$_3$Sn fraction in RRP wires [63]. While this leads to some $J_c$ improvement for small-$d_{sub}$ RRP wires [63], little improvement is seen on high-performance conductors (e.g., those shown in Figure 1) that have fine-grain Nb$_3$Sn fraction of ~60% in subelements [63]. Efforts were also made on improving Nb$_3$Sn fraction by adjusting LAR; however, no improvement is seen relative to the standard wires (e.g., those shown in Figure 1) [64]. These attempts demonstrate that it is difficult to further remarkably improve Nb$_3$Sn fraction in RRP wires.

On the other hand, the fine-grain Nb$_3$Sn fractions in PIT and TT subelements are only 40-45%, so there is still plenty of room for improvement. One way to do so is to decrease the coarse-grain Nb$_3$Sn fraction by increasing the Cu/Sn ratio, as shown in Figure 9. However, it will be seen in the next section that higher Cu/Sn ratio would cause negative impact on the stoichiometry and $B_{c2}$ of Nb$_3$Sn phase. The other way to increase Nb$_3$Sn fraction in PIT and TT wires is to reduce the unreacted Nb fraction under the permission of allowable RRR.

As can be seen from Figure 6, after heat treatment the remaining Nb in a RRP subelement is more or less a uniform layer protecting the Cu stabilizer effectively, while in a PIT and TT subelement a thick Nb barrier must be kept to prevent Sn leakage. What causes this difference? RRP subelements benefit from the Cu network among the Nb filaments. These Cu channels provide shortcuts for Sn diffusion because the diffusion rate of Sn in Cu-Sn is much higher than the Nb$_3$Sn layer growth rate; with a proper ratio of Nb barrier thickness to filament diameter (e.g., ≥2), the filament-packing area can be fully reacted, with part of the Nb barrier left unreacted to protect the Cu stabilizer. On the other hand, a PIT or TT subelement simply uses a Nb alloy tube which has no such Cu network – comparing RRP and TT subelements (Figures 3 b and d, respectively), TT can be regarded as a special distributed-barrier internal-tin type with LAR = 0. In this case, Nb$_3$Sn layer grows outwards as a Sn source/Nb$_3$Sn/Nb diffusion reaction couple in the cylindrical geometry. If sited within a hexagonal Nb alloy tube, this cylindrical Nb$_3$Sn layer can touch the subelement edges and release Sn to the outer Cu matrix while a lot of Nb is still left in the corners. Thus, using round subelements (e.g., Figure 3 c) helps in reducing the unreacted Nb fraction in PIT and TT strands. In practice the situation is more complicated because subelements can become distorted and eccentric after many courses of wire drawing. This is to a large extent tolerable in RRP subelements thanks to the Cu networks, as long as the Nb barriers are not locally thinned. In PIT and TT subelements, however, this is a major problem because the Nb layer can be reacted through on one side while the Nb layer is still thick elsewhere. Thus, the quality of a wire (e.g., regularity, homogeneity along the wire length, etc.), which highly depends on the quality of the starting materials and the processing technology of a manufacturer, can significantly influence the trade-off between the remaining Nb area fraction and wire RRR.

Another possible means to reduce the barrier area fraction without sacrificing RRR is to use materials that react with Sn more slowly than Nb (or Nb-7.5wt.%Ta) does as the barrier. For example, the reaction rate of Nb-Ta alloy (Nb and Ta have unlimited solid solubility) with Sn decreases as the Ta content increases, making pure Ta an ideal barrier material for single-barrier Nb$_3$Sn conductors. However, Ta has much higher work hardening rate and poor ductility compared with Nb, causing breakage issues at small subelement size. On the other hand, the commercial Ta-40wt.%Nb has intermediate ductility and reaction rate with Sn, making itself a promising barrier material [65]. Vanadium was also tried, but it was suspected to degrade $J_c$ [65].

## 5. A theory for Nb$_3$Sn stoichiometry

For Nb$_3$Sn a high $B_{c2}$ requires both proper doping and high Sn content in Nb$_3$Sn. Doping to Nb$_3$Sn can be controlled via the additions to the precursors (e.g., Nb-7.5wt.%Ta, Nb-47wt.%Ti, or Sn-Ti), which is mature for present conductors. On the other hand, high Sn content typically requires high heat treatment temperature (e.g., Figure 5) which is undesirable as grain size increases exponentially with reaction temperature [55]. Evidently a fundamental understanding of what controls Nb$_3$Sn stoichiometry is critical. Based on the studies in the past several decades, several factors have been found to strongly influence the Sn content of Nb$_3$Sn conductors, which are summarized below.

The Sn source can significantly influence the stoichiometry, especially the Sn content gradient, of the Nb$_3$Sn layer. For example, bronze-process strands, of which the Sn sources are bronze with Sn concentration below 9 at.%, tend to have larger Sn content gradients in the Nb$_3$Sn layers than those of RRP, PIT and TT strands, of which the Sn sources are much more Sn-rich (the compositions of Sn sources for different types of Nb$_3$Sn wires are shown in Figure 7). Examples of Sn concentration profiles in Nb$_3$Sn layers of bronze-process and distributed-barrier internal-tin strands are shown in Figure 10.

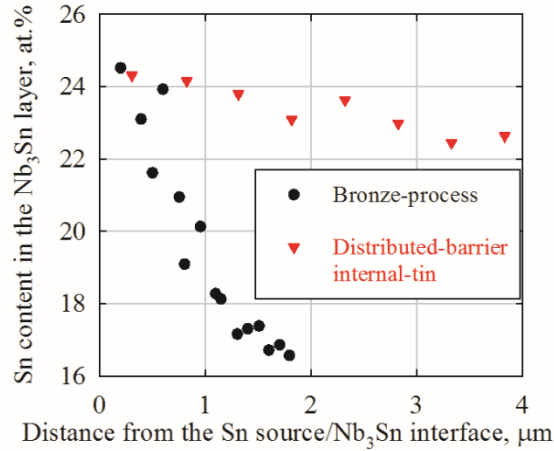

Figure 10. Sn content profiles measured by Energy Dispersive Spectroscopy (EDS)/Scanning Transmission Electron Microscopy (STEM) on Nb$_3$Sn layers in bronze-process [66] and distributed-barrier internal-tin [67] wires.

It can be seen that for both wires the Nb$_3$Sn layers adjacent to the Sn sources have Sn contents above 24 at.%; however, they have very different Sn concentration gradients as the Nb$_3$Sn layers grow thicker. The bronze-process wires have Sn content gradients above 3 at.%/μm [66]. For distributed-barrier internal-tin strands in which original Nb filaments in each subelement merge to a single Nb$_3$Sn annulus after reaction, the Sn content gradient within each original filament may be as high as 0.4-0.5 at.%/μm (Figure 10) [67], although the Sn content gradients from the inner-row filaments to the outer-row filaments are small thanks to the Cu paths for Sn diffusion [68]. The Sn content gradients of TT and PIT subelements are typically 0.1-0.2 at.%/μm [58,69,70]. Such a difference in the Sn content gradients leads to different grain morphologies and superconducting properties (e.g., the $B_{c2}$ values shown in Figure 5). It is found that Nb$_3$Sn layers with larger Sn content gradients tend to have grains of larger aspect ratios. For instance, bronze-process wires are well known to have columnar grains, while grains of RRP, PIT, and TT wires are more or less equiaxed. RRP wires reacted at very low temperatures (e.g., 615 °C) can also have grains with somewhat large aspect ratios [55]. Measurements indicate that boundaries of grains with larger aspect ratios have lower pinning efficiency than those of equiaxed grains [55,68], which is an important reason for the inferior $J_c$s of bronze-process and single-barrier internal-tin wires. Sn sources also influence Nb$_3$Sn layer growth rate: high-Sn Sn sources lead to higher Nb$_3$Sn layer growth rates [71].

Heat treatment temperature is well known to strongly influence the $B_{c2}$, with higher reaction temperature leading to higher Sn content and thus higher $B_{c2}$ (Figure 5). In addition, many studies have demonstrated that $B_{c2}$ increases with reaction extents. For example, Fischer's study on PIT strands showed that highly under-reacted samples have $B_{c2}$s that are only 70-80% of those fully-reacted ones [72]. Such a phenomenon was also observed in TT strands. This could be due to decrease in thermal-strain (induced by the mismatch in thermal contraction coefficients of different components in the composite wire as it is cooled down) with increase of Nb$_3$Sn layer thickness. Moreover, some other factors have also been found to influence Sn content. For example, it was found that Ti doping tends to make Sn content more uniform across the Nb$_3$Sn layer [73], which could explain its positive effect on $B_{c2}$ (Figure 5).

Despite the above empirical conclusions drawn from experiment observations, there still lacks a theory that can explain these facts and illustrate what essentially determines Nb$_3$Sn stoichiometry in conductors. Recently a model has been developed to explore what essentially controls the Sn content of Nb$_3$Sn in a wire [74]. In a subelement, Nb$_3$Sn layer grows in a Sn source/Nb$_3$Sn/Nb diffusion reaction couple at the heat treatment temperature, with a schematic of the system for the planar geometry shown in Figure 11 (a), and the chemical potential of Sn, $\mu_{Sn}$, of the system shown in Figure 11 (b). The Sn sources for Nb$_3$Sn layer growth include Cu-Sn alloys and Nb$_6$Sn$_5$. It is known that Sn is the primary diffusing species in the Nb$_3$Sn phase [75].

The minimum and maximum mole fractions of Sn in $Nb_3Sn$ from the phase diagram are denoted as $X_l$ and $X_u$, respectively (Figure 7), which are around ~0.17 and ~0.26 [76]. The Sn source/$Nb_3Sn$ and $Nb_3Sn$/Nb interfaces are noted as Interfaces I and II, respectively.

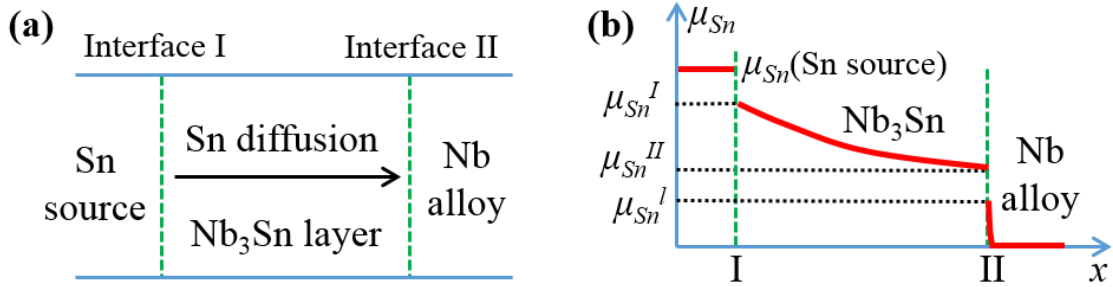

Figure 11. Schematics of (a) a Sn source/$Nb_3Sn$/Nb diffusion reaction system in the planar geometry, and (b) the $\mu_{Sn}$ profile of the system, where $\mu_{Sn}^I$ and $\mu_{Sn}^{II}$ stand for $\mu_{Sn}$s of $Nb_3Sn$ at the Interfaces I and II, respectively, and $\mu_{Sn}^l$ stands for $\mu_{Sn}$(Nb-$X_l$Sn).

The assumptions and derivations for the governing equations, which are based on Fick's laws and mass conservation, for this diffusion reaction system are described in details in [74]. Theoretically the model can quantitatively calculate the chemical potential, composition profile, and growth rate of a growing $Nb_3Sn$ layer by solving the governing equations. Here this section focuses on analyzing the $Nb_3Sn$ layer stoichiometry qualitatively.

Before discussing the composition of $Nb_3Sn$, it must be ascertained whether Sn diffuses through the $Nb_3Sn$ layer via bulk diffusion or grain boundary diffusion. In the case of bulk diffusion, for a Cu-Sn/$Nb_3Sn$/Nb system, as Sn content and $\mu_{Sn}$ of Cu-Sn drop with the growth of $Nb_3Sn$ layer, $X_{Sn}$ of the entire $Nb_3Sn$ layer should decrease with $\mu_{Sn}$(Cu-Sn), because $\mu_{Sn}$(Sn source) $\geq \mu_{Sn}$($Nb_3Sn$) $\geq \mu_{Sn}$(Nb- $X_l$ Sn). Finally, one of the following two cases would occur: (1)

if Nb is in excess in the Sn source/Nb$_3$Sn/Nb system, $\mu_{Sn}$(Sn source) eventually drops to $\mu_{Sn}$(Nb-$X_l$ Sn), so the system ends up with the equilibrium among Nb, Nb- $X_l$ Sn, and low-Sn Cu-Sn (as shown by the shaded region in the Cu-Nb-Sn phase diagram in Figure 7), in which case the Sn content in the whole Nb$_3$Sn layer would be $X_l$; (2) if the Sn source is in excess, then Nb would eventually be consumed up and Nb$_3$Sn gets homogenized with time and finally $\mu_{Sn}$(Nb$_3$Sn) = $\mu_{Sn}$(Sn source). In either case, Nb$_3$Sn layer stoichiometry eventually reaches homogeneity. However, it is found that the real characteristics for Nb$_3$Sn compositions are different from the above bulk diffusion case. For example, in PIT and TT subelements in which Nb is in excess, even after extended annealing times after the Nb$_3$Sn layers have finished growing (which indicates that the Sn sources have been depleted, i.e., $\mu_{Sn}$s have dropped to $\mu_{Sn}^l$), $X_{Sn}$s of Nb$_3$Sn remain high above $X_l$, without dropping with annealing time. For example, as the reaction time of a PIT strand was extended from 64 hours (at which time the Sn source had been depleted) to 768 hours at 675 °C, the Sn content of Nb$_3$Sn remained above 22 at.% [54], far above the $X_l$ of ~17 at.%. Moreover, the Sn content gradients in the Nb$_3$Sn layers remain after extended annealing.

The above peculiarities indicate that bulk diffusion in Nb$_3$Sn is frozen and grain boundary diffusion dominates. In fact, it was reported in [77] that the bulk diffusivity of Sn in Nb$_3$Sn is lower than $10^{-23}$ m$^2$/s at 650 °C. Using the relation $l \approx \sqrt{Dt}$, it can be estimated that the bulk diffusion distance is less than ~1 nm for a duration of 100 hours at 650 °C. Instead, the diffusivity of Sn along Nb$_3$Sn grain boundaries is several orders of magnitudes higher than that in the bulk [77], which, in combination with the small Nb$_3$Sn grain size (~100 nm), leads to the domination of grain boundary diffusion. A schematic of the diffusion reaction process for grain boundary diffusion is shown in Figure 12.

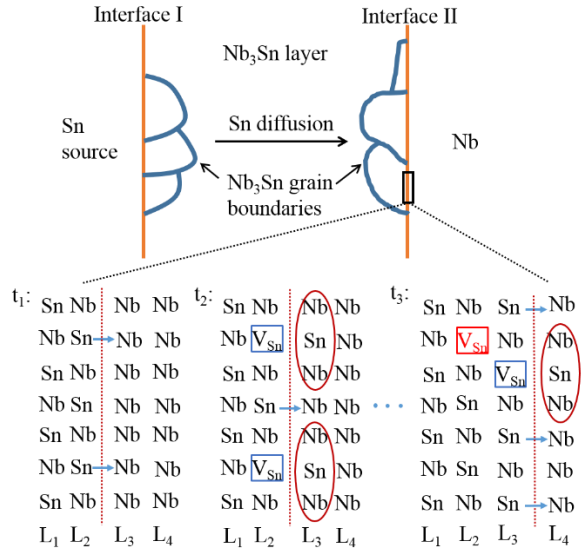

Figure 12. A schematic of the diffusion reaction process for grain boundary diffusion.

At the Nb$_3$Sn/Nb interface, high-Sn Nb$_3$Sn (L$_2$ layer in Figure 12) reacts with Nb (L$_3$ layer) to form some new Nb$_3$Sn cells, leaving Sn vacancies (noted as V$_{Sn}$s) in the L$_2$ layer (time $t_2$). If bulk diffusivity is high, V$_{Sn}$s simply diffuse through bulk (e.g., from L$_2$ to L$_1$) to the Sn source. If bulk diffusion is frozen, the V$_{Sn}$s diffuse first along Nb$_3$Sn/Nb inter-phase interface and then along Nb$_3$Sn grain boundaries to the Sn source. This process continues until this L$_3$ layer entirely becomes Nb$_3$Sn (time $t_3$), so the reaction frontier moves ahead to L$_3$/L$_4$ interface, while the L$_2$/L$_3$ interface now becomes an inter-plane inside the Nb$_3$Sn lattice. If bulk diffusion is completely frozen, the V$_{Sn}$s in the L$_2$ layer that fail to diffuse into the Sn source before the Nb$_3$Sn/Nb inter-phase interface moves ahead from the L$_2$/L$_3$ interface will be frozen in this L$_2$ layer forever, and will perhaps transform to Nb-on-Sn anti-site defects later (for Nb$_3$Sn, Nb-on-Sn anti-sites are more stable than Sn vacancies [78]). Since these point defects determine the Nb$_3$Sn composition, the $X_{Sn}$ in this L$_2$ layer cannot change anymore regardless of $\mu_{Sn}$ variations in grain boundaries. That is to say, $X_{Sn}$ of any point in a Nb$_3$Sn layer is just the Sn content of the moment when the reaction frontier sweeps across this point, i.e., the $X_{Sn}(x)$ of the Nb$_3$Sn layer is simply an accumulation of Sn contents at Interface II with layer growth. This explains why Nb$_3$Sn layer composition cannot reach homogeneity.

Then what determines the Nb$_3$Sn composition? From Figure 12 it can be clearly seen that there is a competition between two processes which determines the V$_{Sn}$ fraction in the frontier Nb$_3$Sn layer: the reaction at the Nb$_3$Sn/Nb interface produces V$_{Sn}$s in the L$_2$ layer, while the diffusion of Sn along Nb$_3$Sn grain boundaries fills these V$_{Sn}$s. Thus, if the diffusion rate is slow relative to the reaction rate at Interface II, a high fraction of V$_{Sn}$s would be left behind as the Interface II moves ahead, causing low Sn content. If, on the other hand, the diffusion rate is high relative to the reaction rate at Interface II, the Nb$_3$Sn layer has enough time to get homogenized with the Sn source, causing low $X_{Sn}$ gradient. Thus, higher diffusion rate relative to reaction rate benefits Nb$_3$Sn stoichiometry.

Next let us consider the influence of Sn source on Nb$_3$Sn composition. In order to do this, the relation between $\mu_{Sn}(X_{Sn})$ of Cu-Sn and $\mu_{Sn}(X_{Sn})$ of Nb$_3$Sn is required. The Cu-Sn system has been well studied, and the phase diagram calculated by the CALPHAD technique using the thermodynamic parameters given by [79] is well consistent with the experimentally measured Cu-Sn phase diagram [80]. Thus, the parameters from [79] are used to calculate the $\mu_{Sn}$ of Cu-Sn, which is shown in Figure 13. On the other hand, although thermodynamic data of Nb-Sn system were proposed by [79] and [81], in these studies Nb$_3$Sn was treated as a line compound, making their data unreliable. However, some information about $\mu_{Sn}$ of Nb$_3$Sn can be inferred from its relation with $\mu_{Sn}$ of Cu-Sn: since in bronze-process wires Cu-7 at.% Sn leads to the formation of Nb-24 at.% Sn near the Cu-Sn source [66], we have $\mu_{Sn}$(Cu-7 at.% Sn) ≥ $\mu_{Sn}$(Nb-24 at.% Sn).

Thus, the $\mu_{Sn}(X_{Sn})$ curve for Nb$_3$Sn must be concave up and is approximately shown in Figure 13. Furthermore, it can also be inferred that the Sn transfer rate at the Cu-Sn/Nb$_3$Sn interface must be much faster than that at the Nb$_3$Sn/Nb interface, so $\mu_{Sn}$ discontinuity across the Interface I is small.

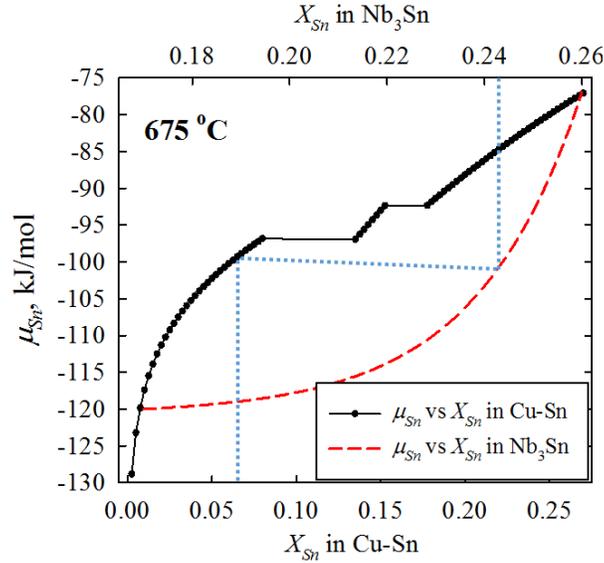

Figure 13. The $\mu_{Sn}(X_{Sn})$ curve for Cu-Sn calculated from the thermodynamic data given in [79], and a rough $\mu_{Sn}(X_{Sn})$ curve for Nb$_3$Sn speculated based on the phase formation relation between Cu-Sn and Nb$_3$Sn. The blue dotted lines show how Cu-7 at.%Sn can form Nb$_3$Sn with Sn content above 24 at.% Sn at the bronze/Nb$_3$Sn interface. Note that the approximate $\mu_{Sn}(X_{Sn})$ curve for Nb$_3$Sn is only for showing the trend.

The phenomenon that low-Sn Sn sources lead to larger $X_{Sn}$ gradients and lower growth rates of Nb$_3$Sn layers also originates from the concave-up $\mu_{Sn}$-$X_{Sn}$ curve for Nb$_3$Sn. Schematics for the $\mu_{Sn}$ profiles of the Sn source/Nb$_3$Sn/Nb system for two different Sn sources are shown in Figure 14 (a). Based on the calculations using the model in [74], it is known that larger $\mu_{Sn}$ in the Sn source causes both larger $\mu_{Sn}$ drop across the Nb$_3$Sn layer and larger ($\mu_{Sn}^{II}-\mu_{Sn}^{I}$), which cause larger diffusion rate and larger reaction rate, respectively. This explains the fact that increase in the activity of Sn in the Sn source (e.g., increasing Sn content of Cu-Sn) leads to higher Nb$_3$Sn layer growth rate. Figure 14 (b) shows a schematic of the Nb$_3$Sn $X_{Sn}(\mu_{Sn})$ plot transformed from Figure 13, and the $X_{Sn}$ drop, $\Delta X_{Sn}$, caused by the $\mu_{Sn}$ drop, $\Delta\mu_{Sn}$, across the Nb$_3$Sn layer for the two cases shown in Figure 14 (a). As can be seen, although higher $\mu_{Sn}$(Sn source) causes larger $\mu_{Sn}$ drop across the Nb$_3$Sn layer, $X_{Sn}$ drop is smaller due to the curved $X_{Sn}(\mu_{Sn})$ of Nb$_3$Sn, explaining why strands with high-$\mu_{Sn}$ Sn sources (such as RRP, PIT, and TT) have smaller Sn at.% gradients than those with low-$\mu_{Sn}$ Sn sources (such as bronze-process and single-barrier internal-tin).

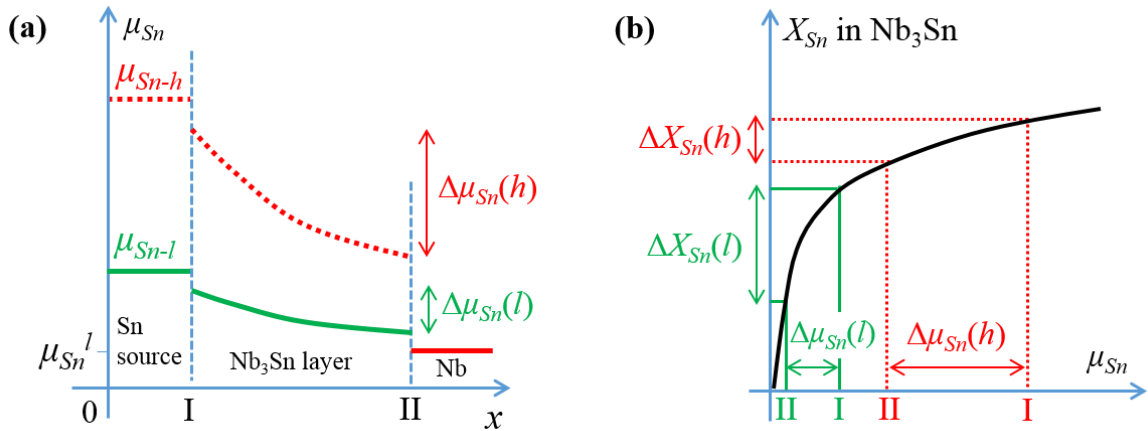

Figure 14. (a) Schematics of the $\mu_{Sn}$ profiles of the system for two different Sn sources – high $\mu_{Sn}$ (denoted as $\mu_{Sn-h}$) and low $\mu_{Sn}$ (denoted as $\mu_{Sn-l}$), and (b) schematic of the $\Delta X_{Sn}$ caused by the $\Delta\mu_{Sn}$ across the Nb$_3$Sn layer. Note in (b) how a small $\Delta\mu_{Sn}$ across the Nb$_3$Sn layer in the low-$\mu_{Sn}$ region can cause a large $\Delta X_{Sn}$.

PIT and TT strands have Sn sources with higher $\mu_{Sn}$s than RRP strands (Figure 7), so a large amount of $Nb_6Sn_5$ is formed in PIT and TT strands, and such $Nb_6Sn_5$ serves as the Sn source before Cu-Sn alloys (which have lower $\mu_{Sn}$s than $Nb_6Sn_5$ does) takes over, which explains why $X_{Sn}$ gradients across the $Nb_3Sn$ layers in PIT and TT strands are low.

As shown in Section 4, higher Cu/Sn ratio in RRP, PIT and TT strands leads to higher fine-grain $Nb_3Sn$ fraction due to reduction in amounts of Nb-Sn intermediate phases and coarse-grain $Nb_3Sn$. Could this be used to improve $J_c$? The non-matrix $J_c$s of the three TT wires displayed in Figure 8 are shown in Figure 15. As Cu/Sn ratio increases, although fine-grain $Nb_3Sn$ fraction does increase, the $J_c$s drop dramatically. This is mainly due to the decrease of $B_{c2}$ and $F_{p,max}$. This is consistent with observations in internal-tin strands of which layer $J_c$ increases as Cu/Sn ratio decreases [68], as mentioned in Section 2. The 4.2 K $B_{c2}$ values of the three TT wires obtained from the Kramer plots are 24, 22, and 21 T as Cu/Sn ratio increases, respectively. It is also observed that the $Nb_3Sn$ layer growth rate is the highest for the wire with the lowest Cu/Sn ratio.

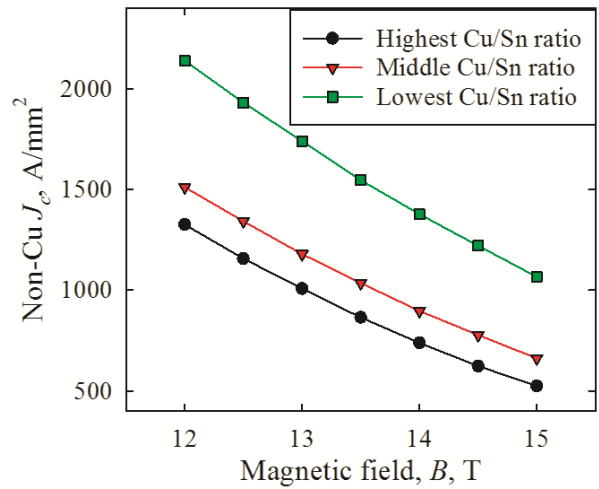

Figure 15. The non-matrix $J_c$s (4.2 K) of the three TT wires shown in Figure 8.

Figure 16 shows SEM images on fractured surfaces of the three wires: columnar grains can be seen in each wire at the $Nb_3Sn/Nb$ interface. Such columnar grains can be observed in RRP and PIT wires as well. This can be explained by the theory in this section: in the last stages of $Nb_3Sn$ layer growth, the Cu-Sn source has low Sn content and $\mu_{Sn}$, causing larger $X_{Sn}$ gradient in the growing $Nb_3Sn$ layer. From Figure 16 it is clear that as Cu/Sn ratio increases, the columnar-grain layer becomes thicker. These columnar-grain $Nb_3Sn$ layers have large $X_{Sn}$ gradients (which is demonstrated by EDS results [58]), explaining the low $B_{c2}$s for the wires with high Cu/Sn ratios. Moreover, because boundaries of columnar grains have low pinning efficiency, a larger fraction of such columnar grains leads to lower $F_{p,max}$.

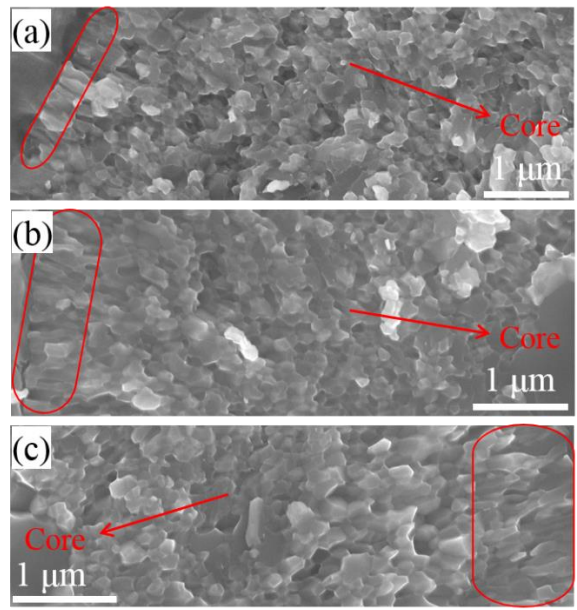

Figure 16. SEM images of the TT wires with (a) low, (b) medium and (c) high Cu/Sn ratios, with the columnar-grain layers marked.

It is interesting to note that RRP wires with similar Cu/Sn ratio to that of the TT wire with the highest Cu/Sn ratio in Figure 15 have considerably higher $B_{c2}$. This is because in RRP wires, thanks to the Cu paths that transport Sn among Nb filaments, the Sn source/Nb$_3$Sn/Nb diffusion reaction occurs with filaments as units; as a result, the Sn diffusion distance in a Sn source/Nb$_3$Sn/Nb unit in RRP wires is less than 1 μm (the radius of a filament), while those of PIT and TT wires are larger than 5 μm. It is clear from the model developed in [74] that short diffusion distance helps in reducing $X_{Sn}$ inhomogeneity in the Nb$_3$Sn layer.

As another example, Zn addition to the Cu-Sn source was found to enhance the Nb$_3$Sn layer growth rate and the Sn content of the Nb$_3$Sn layer as well [82]. This effect can also be explained by the above model, given that Zn addition to Cu-Sn increases the activity of Sn in the Cu-Sn source [83].

Heat treatment temperature can simultaneously influence multiple factors, such as chemical potential of Sn in the Sn source, diffusion rate of Sn in Nb$_3$Sn, and reaction rates at both interfaces, etc. Moreover, from the thermodynamic point of view, the $\mu_{Sn}(X_{Sn})$ of both Cu-Sn and Nb$_3$Sn and their relation may also change with temperature. Thus, to explain the influence of heat treatment temperature on the Sn content of Nb$_3$Sn, it needs studies about change of each of these factors with temperature.

Another factor that may influence the Sn content of Nb$_3$Sn, according to the above theory, is the competition between the Sn diffusion rate across the Nb$_3$Sn layer and the Nb-Sn reaction rate at the Nb$_3$Sn/Nb interface. This effect has hardly been noticed; however, the phenomenon that Ti addition tends to accelerate Nb$_3$Sn layer growth rate [55] and make the stoichiometry more uniform in the Nb$_3$Sn layer [73] could be explained by this effect. It has been seen that Ti addition segregates at Nb$_3$Sn grain boundaries and causes greater lattice distortion at grain boundaries, which is believed to cause higher grain boundary diffusivity [84]. According to the theory presented in this section, higher grain boundary diffusion rate leads to not only higher Nb$_3$Sn layer growth rate, but also reduced $X_{Sn}$ gradient, because higher Sn diffusion rate relative to Nb-Sn reaction rate benefits Sn content homogenization.

In summary, according to the theory presented in [74] and in this section, the Sn content of Nb$_3$Sn layer can be improved by: a) increasing the $\mu_{Sn}$ of the Sn source, b) increasing the diffusivity of Sn along the Nb$_3$Sn grain boundaries, c) reducing Nb$_3$Sn grain size which leads to more grain boundaries that can transport Sn faster to the reaction frontier, d) decreasing the diffusion distance by reducing the subelement (filament) size, (e) increasing reaction rate at the Sn source/Nb$_3$Sn interface to decrease the $\mu_{Sn}$ drop across it, and f) decreasing reaction rate at the Nb$_3$Sn/Nb interface to increase the $\mu_{Sn}$ drop across it.

## 6. Prospects to improve pinning capacity

The average grain sizes of present Nb$_3$Sn strands (110-150 nm) are ten times larger than the flux line spacing at 12-20 T, which is 10-15 nm. In this case, a large fraction of flux lines are not directly pinned by grain boundaries; the pinning force for such flux lines is provided by the rigidity of the flux line lattice (*i.e.*, the repulsion force among flux lines, which transfers the Lorentz force on each flux line to the leading flux lines that are directly pinned by grain boundaries). Such pinning force is limited by the shear strength of the flux line lattice, which is weak relative to the pinning force provided directly by grain boundaries. In this case, the $F_p(B)$ curves peak at ~0.2$B_{c2}$ [52]. The pinning efficiency can be significantly improved if grain size is refined. Many studies have shown that $F_{p,max}$ is inversely proportional to grain size [37,38,55,72].

Furthermore, experiments on Nb$_3$Sn thin films [85] showed that if Nb$_3$Sn grain sizes were reduced to 15-30 nm, the peaks of their $F_p(B)$ curves would shift to $0.5B_{c2}$. As grain size is comparable to flux line spacing, flux lines interact with grain boundaries individually, so loss in superconductivity is possible only if the Lorentz force is high enough for flux lines to break free from the direct pinning by grain boundaries; in this case $F_p$ is a summation of pinning force on each flux line, making the $F_p(B)$ curve peak at $0.5B_{c2}$. This "individual pinning" mode has been realized in Nb-Ti superconductors, in which non-superconducting α-Ti precipitates with space below 10 nm serve as the primary flux pinning centers.

The shift of the $F_p(B)$ curve peak from $0.2B_{c2}$ to $0.5B_{c2}$ brings not only significant improvement in high-field $J_c$ but also remarkable reduction of low-field $J_c$ and associated magnetization, because the $J_c(B)$ curve is linear as the $F_p(B)$ curve peaks at $0.5B_{c2}$ – by contrast, as the $F_p(B)$ curve peaks at $0.2B_{c2}$, $J_c$ increases sharply as field decreases at low fields (≤4 T). Since magnetization is the driving force for low-field flux jumps and field errors in magnets, decreasing the low-field $J_c$ by shifting the $F_p(B)$ curve peak to $0.5B_{c2}$ also benefits in improving low-field stability and suppressing the field errors caused by persistent-current magnetization [10,86].

The primary method used at present to reduce Nb$_3$Sn grain size is to decrease heat treatment temperature for Nb$_3$Sn formation. However, even reducing the temperature to 615 °C, the grain size is still 90-110 nm [55], and decreasing reaction temperature is adverse to Nb$_3$Sn stoichiometry and $B_{c2}$. Another means to increase pinning is introducing additional pinning centers (APCs) to Nb$_3$Sn wires, which has been heavily pursued since the 1980s. Flukiger et al. tried to use Nb containing thin Ta ribbons for wire fabrication, expecting the normal Ta metal to serve as APCs [87]. Some improvements were seen; however, this was believed to be caused by increased $B_{c2}$ (due to Ta doping) rather than increase of pinning [88]. Later Cu ribbons were used instead of Ta, and this idea of introducing Cu ribbons to Nb alloy precursors was actively pursued by a few groups for two decades [89-92]. In this method, filaments composed of Cu-compassed Nb (with Cu and Nb thicknesses of 20-40 and 30-100 nm, respectively) were produced by successive steps of bundling and drawing of Cu-Nb composites [91]. These non-superconducting Cu ribbons (which transformed to Cu-Sn at the reaction stage) were expected to serve as APCs impeding flux line motion. However, this approach failed to produce conductors with $J_c$ higher than regular $Nb_3Sn$ wires [90,92]. A possible reason is that after heat treatment the Cu-Sn became isolated, widely-separated voids [92], which could not truly serve as APCs. In fact, one should remember that the Cu networks among Nb filaments in RRP subelements, which are even thicker (hundreds of nanometers), still disappear after heat treatment. These efforts demonstrate that introducing nano metal ribbons to Nb alloy precursors as APCs is not likely feasible.

Efforts were also made to introduce second phase particles to Nb precursors. One such effort was made by Motowidlo to add the rare earth elements Y or Gd to Nb melt to make Nb-Y or Nb-Gd alloy [93]. Because Y or Gd has little solubility in Nb, it forms fine precipitates in Nb. Motowidlo's experiments showed that the addition of 0.75 at.% Y led to a refinement of the average grain size from 400-500 nm to 200-300 nm for samples reacted at 750 °C [93]. Magnetization measurements showed that the $F_{p,max}$ of the Y-added sample was nearly twice that of the control sample [93]. Perhaps a larger amount of Y addition can lead to an even more significant reduction of grain size. Nevertheless, a problem with this method is, even this small amount of addition hardened the Nb alloy enough to make wire processing difficult [93].

Indeed, the requirement for fine filament size makes introduction of APCs to $Nb_3Sn$ wires a challenging job. To obtain small $d_{sub}$, manufacturing of $Nb_3Sn$ strands requires many courses of extrusion/drawing to attain the final wire size, with stretch ratios of Nb alloy precursors typically above 100,000. Hence, the precursors must be very ductile. This precludes the possibility of dispersing particles in the starting Nb alloy, as precipitation hardening makes the subsequent reduction difficult. Then there is only one possible means to introduce APCs into $Nb_3Sn$ wires, which is to form such APCs through chemical reactions during heat treatment. However, it is impossible to find a Nb-M alloy such that the metal M reacts with Sn to form compound precipitates in $Nb_3Sn$. Thus, it requires introduction of an additional element (such as oxygen) that can react with M to form compound precipitates: a feasible approach to realize this is the internal oxidation method, which forms oxide compound of M.

Internal oxidation of an A-B solid solution means that as oxygen is supplied to it, only the solute B is selectively oxidized. The oxide product of the solute may precipitate out in the matrix (solvent A) in the form of fine particles, which could be used for dispersion strengthening or grain refinement. To make the internal oxidation possible, the solute B must be much less noble than the solvent A. The internal oxidation method was successfully used in $Nb_3Sn$ tapes in the 1960s [94]. Commercially available Nb-1wt.%Zr alloy was used because Zr has much stronger affinity to oxygen than Nb does. The Ellingham diagram (which shows the formation energies of oxides) of some metals is shown in Figure 17 [95]. Nb-1%Zr foil was anodized to form $Nb_2O_5$ layer on the surface, which decomposed during the subsequent annealing and released oxygen to Nb-1%Zr. The foil was then coated with Cu-Sn and reacted at 1050 °C to form $Nb_3Sn$ containing $ZrO_2$ particles [96]. With sufficient oxygen supplied, the grain size was refined to 300 nm, compared with 2-3 µm for samples without any oxygen addition [96].

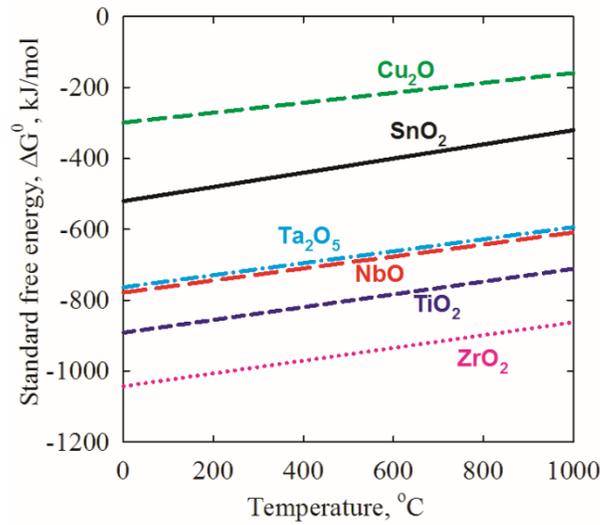

Figure 17. The formation energies of some metal oxides [95].

After Nb$_3$Sn tapes were replaced by wires in the early 1970s, a lot of efforts have been made to transfer the internal oxidation method to Nb$_3$Sn wires. This turned out to be quite challenging. In a Nb$_3$Sn wire, due to the presence of the outside Cu sheath, oxygen cannot be delivered to the Nb-Zr alloy externally (e.g., by anodization or by annealing in an oxygen atmosphere). Benz ([97]) and Zeitlin [98] tried to pre-dissolve oxygen in Nb-1%Zr alloy and to use this oxygen-containing Nb-1%Zr for wire fabrication. However, the oxygen-containing Nb-1%Zr alloy was hard to draw, because the pre-dissolved oxygen markedly increased its strength and decreased its ductility. Subsequently, Zeitlin tried to use oxide powder as oxygen source in a mono-element-internal-tin (MEIT) wire, with the core composed of a mixture of SnO$_2$ and Sn powders [98,99]. However, no noticeable grain refinement was observed [98,99].

The work of internal oxidation was restarted recently [100,101]. Zeitlin's work was analyzed with the aim of finding out why the grains of the wires were not refined as those of the Nb$_3$Sn tapes were. It was concluded that the Nb-Zr alloy in that wire did not really obtain the oxygen in the $SnO_2$ powder [100]. Then a series of calculations and experiments were done to reveal the characteristics of oxygen transfer from oxide powder to Nb alloy [59,100]. They demonstrate that direct contact between oxide powder and Nb alloy is not necessary for oxygen transfer to occur – as long as the atmosphere connects, oxide powders can supply oxygen to the Nb alloy through atmosphere [102]; on the other hand, the oxygen diffusion rate in Cu is too slow to supply sufficient oxygen to Nb-1%Zr, unless the Cu layer is very thin [59]. Thus, the subelement structure must be properly designed to avoid cases in which inert Cu layer blocks the path of oxygen transfer.

Based on these findings about oxygen transfer, the subelement structures of various types of $Nb_3Sn$ strands were modified such that oxide powders are located in proper positions inside the subelements to supply sufficient oxygen to oxidize the Zr atoms during heat treatment. A feasible filament structure enabling internal oxidation in a TT wire is illustrated in Figure 18 (a). The filament is comprised of four concentric layers (the central Sn rod, the Cu layer, the oxide powder layer, and the Nb-1%Zr alloy tube) inside the Cu sheath. Monofilaments following this structure using $SnO_2$ or $NbO_2$ powder were fabricated; $NbO_2$ supplies little oxygen to Nb-Zr [100], and was thus used as a control. Both wires were heat treated at 625 °C and 650 °C.

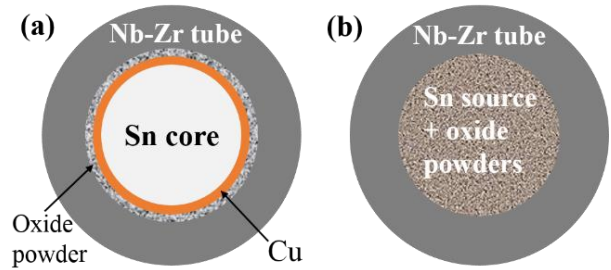

Figure 18. Schematics of modified (a) TT and (b) PIT filaments to realize internal oxidation.

From SEM images of fractured surfaces of the wires reacted at 650 °C (Figure 19), the average grain sizes of the wires with $NbO_2$ and $SnO_2$ were calculated as 104 nm and 43 nm, respectively. The wires with $NbO_2$ and $SnO_2$ reacted at 625 °C had average grain sizes of 81 nm and 36 nm, respectively.

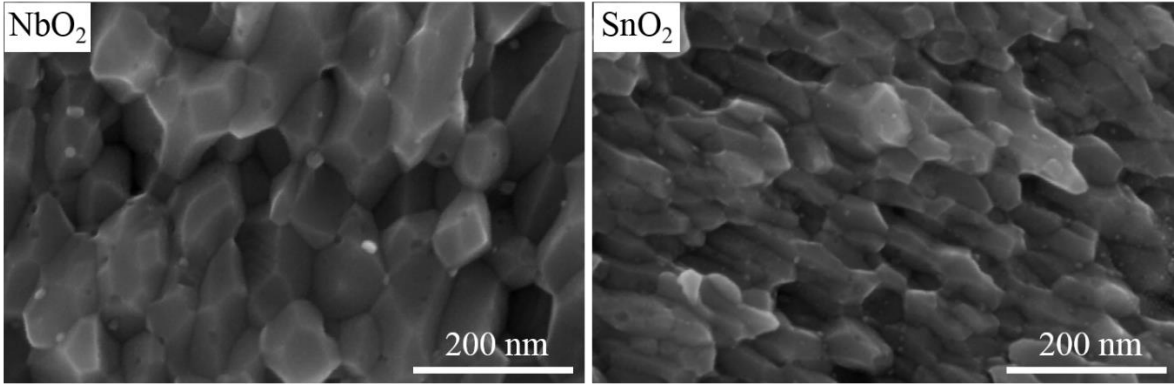

Figure 19. Magnified SEM images on fractured surfaces of the filaments with $NbO_2$ and $SnO_2$ powders, both reacted at 650 °C for 150 hours.

The grain sizes of samples heat treated at various temperatures up to 850 °C show that internal oxidation refines the grain size over the whole temperature range [100]. TEM images show that in the filament with $SnO_2$, there are both intra-granular and inter-granular $ZrO_2$ particles, with sizes ranging from 3 to 15 nm [100].

$Nb_3Sn$ layer $J_c$s were calculated from magnetic moment vs field loops. The 4.2 K, 12 T layer $J_c$s of the wire with $NbO_2$ reacted at 625 °C is 4400 A/mm$^2$ [101], close to those of RRP strands. The value of the wire with $SnO_2$ reacted at 625 °C is 9600 A/mm$^2$ [101]. The $F_p(B)$ curves were calculated from the $J_c(B)$ curves and fitted to a universal scaling law $F_p = C \cdot F_{p,max} \cdot b^p \cdot (1-b)^q$ (where $C$ is a constant, and $b=B/B_{c2}$) [101]. The $Nb_3Sn$ layer $F_{p,max}$ of the wire with $SnO_2$ reacted at 625 °C is 180 GN/m$^3$, doubling those of RRP conductors due to refined grain size. The $B_{c2}$ of the wire with $NbO_2$ reacted at 625 °C is 20.9 T, while those of the wire with $SnO_2$ are 23 T and 20 T for samples reacted at 650 °C and 625 °C [101]. These values are similar to those obtained in regular binary PIT wires reacted at 675 °C [72], indicating that introduction of $ZrO_2$ nanoparticles to $Nb_3Sn$ and refinement of grain size do not affect $B_{c2}$. By normalizing $F_p$ to $F_{p,max}$ and $B$ to $B_{c2}$, normalized $F_p(B)$ curves were generated, which peak at ~$0.2B_{c2}$, $0.26B_{c2}$, and $1/3B_{c2}$ for the wire with $NbO_2$ reacted at 625 °C and the wire with $SnO_2$ reacted at 650 °C and 625 °C, which have grain size of 81 nm, 43 nm, and 36 nm, respectively [101].

What is the mechanism for the shift of $F_p(B)$ curve peak to one third of $B_{c2}$ for the sample with grain size of 36 nm? One possibility is that this is due to point pinning by $ZrO_2$ particles, following the point-pinning model proposed by Dew-Hughes [103]. If this is the case, the $F_p(B)$ curve peak cannot shift beyond $1/3B_{c2}$. An alternative explanation is, as the refined grain size becomes closer to the flux line spacing, some individual pinning begins to occur; if this mechanism is true, then the $F_p(B)$ curve can eventually shift to $0.5B_{c2}$ by further refinement of grain size. To find out which mechanism is correct, it is necessary to further refine the grain size and see if the peak can shift further. To obtain smaller grain size, the above wire was heat treated at 600 °C, which led to an average grain size of ~30 nm (SEM image of a fractured surface is shown in Figure 20 a). A fit to its $F_p(B)$ curve at 4.2 K (Figure 20 b) shows that the curve peaks at ~$0.4B_{c2}$, indicating that the second mechanism is correct. An extrapolation shows that the $F_p(B)$ curve peak can shift to $0.5B_{c2}$ by further refining grain size to roughly 20-25 nm.

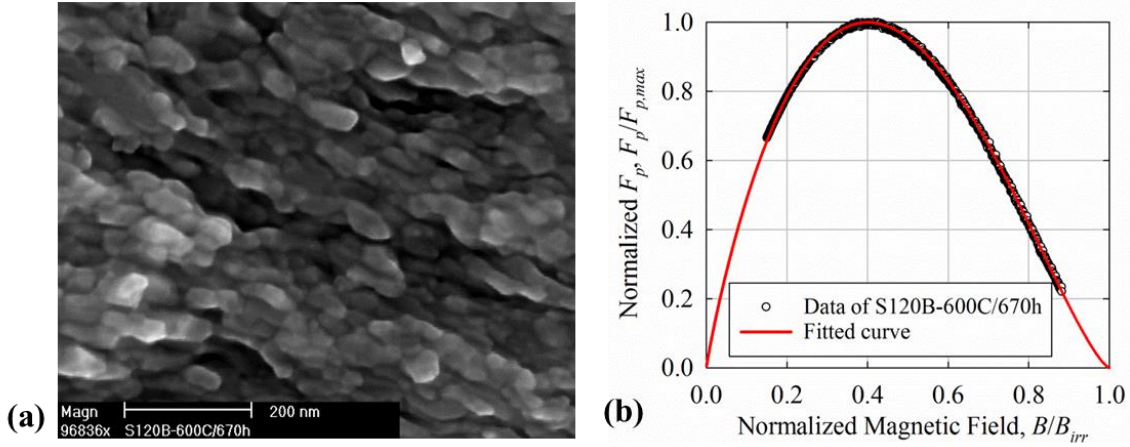

Figure 20. (a) SEM image of the fractured surface of the wire with $SnO_2$ powder reacted at 600 °C, and (b) its normalized $F_p(B)$ curve at 4.2 K with a fit using the general scaling law.

It is concluded from these studies that as grain size is large (e.g., nearly 100 nm), a decrease of the grain size only shifts the $F_p(B)$ curve upward (i.e., $F_{p,max}$ increases) while the peak remains at $0.2B_{c2}$; however, as grain size drops below ~50 nm, not only does $F_{p,max}$ increase, but also the $F_p(B)$ curve peak shifts to higher fraction of $B_{c2}$, which gives extra benefit for high-field $J_c$ improvement [101].

The above results were obtained on a TT subelement based on the scheme shown in Figure 18 (a). Such a structure is difficult to obtain small $d_{sub}$, however. Figure 18 (b) shows a scheme to realize internal oxidation in a PIT filament. The filament is comprised of a Cu-jacketed Nb-Zr alloy tube, which is filled with a mixture of Sn source (e.g., Cu-Sn or Nb-Sn) and oxide powders. This structure turns out to be much more amenable to wire drawing. A PIT wire with 61 filaments produced in HTR was drawn down to 0.254 mm (with $d_{sub}$ of 24 μm) without breakage [102]. During heat treatment, the oxide powder in the core releases oxygen to the Nb-1%Zr. Experiments demonstrated that with proper oxide amount, the Nb-1%Zr alloy was fully oxidized after suitable heat treatments, which led to grain sizes similar to the TT wires shown in

Figure 19. It is worth pointing out that proper intermediate heat treatment schedule is critical for oxygen transfer in PIT wires [102]. Transport (voltage-current) measurements on such PIT wires revealed transport $J_c$s similar to the magnetic layer $J_c$s mentioned earlier [104].

Development of such PIT wires is under way to fulfill the requirements of high non-matrix $J_c$, small $d_{sub}$, high RRR, and large piece length. In order to obtain high non-matrix $J_c$, while we improve pinning by refining grain size, we also need to optimize the other two factors (fine-grain Nb$_3$Sn fraction and $B_{c2}$) so that they are not considerably lower compared with those of regular Nb$_3$Sn conductors. Fortunately, the amount of oxide powder needed to fully oxidize the Nb-1%Zr alloy only occupies several percent of the subelement volume [102]; thus, use of this technique causes very small influence on real estate. This, however, requires optimization of precursor ratios in this new type of conductors, including Nb/Sn, Sn/Cu, and O/Zr, which needs R&D work.

The $B_{c2}$s of the above internally-oxidized wires are not optimized as a result of under-reaction and lack of ternary doping. The wires can be fully reacted in a reasonable time if the $d_{sub}$ is small and the Nb/Sn ratio is suitable. Adding ternary dopants, however, requires special attention. Since oxide powder is preferably added to the core of each subelement for the ease of wire processing, introduction of Ti to the Sn core as in regular Nb$_3$Sn wires would cause a reaction between Ti and O to form TiO$_2$ [102]. One way to solve this problem is to move the Ti source away from the core (e.g., distributing Nb-47%Ti rods among the Nb alloy as in [105]). Another way is to use Ta as dopant, because Ta has similar affinity to oxygen than Nb does (Figure 17). Suppose by adding dopants, the 4.2 K $B_{c2}$ can be increased to 25 T while the grain size is kept at 36 nm, and at such a grain size, the $F_p(B)$ curve still peaks at $1/3B_{c2}$ and $F_{p,max}$ is still 180 GN/m$^3$ [101], then the 15-20 T layer $J_c$s are estimated to be over 3 times of those of present RRP wires [104].

It is expected that the internal oxidation technique can also be implemented in RRP wires by making suitable modifications of subelements. One promising scheme is to use a mixture of Sn source powder and oxide powder to replace the Sn rod. Experiments have shown that oxygen in the core can diffuse through the Cu-Nb-Sn ternary phase [102]. A possible scheme for applying this technique to single-barrier internal-tin strands is to use Nb-Zr tubes filled with oxide powder to replace Nb rods – however, it needs verification whether desirable filament sizes can be obtained in this structure.

## 7. Prospects for further improvement of Nb$_3$Sn superconductors

It is shown in Section 3 that the non-matrix $J_c$ of Nb$_3$Sn strands is determined by three factors: current-carrying Nb$_3$Sn fraction in subelements, $B_{c2}$, and pinning capacity, with the latter two determining the layer $J_c$. Table II summaries the three factors of state-of-the-art RRP, PIT and TT strands, and the presumed limits or optimal levels that Nb$_3$Sn conductors can possibly reach. Based on an analytic model developed for the variation of phase fractions with precursor ratios, it is shown in Section 4 that the maximum current-carrying Nb$_3$Sn fraction in a subelement can reach ~65%. Nb$_3$Sn strands with optimized doping and very high heat treatment temperature have $B_{c2}$s (4.2 K) of ~28 T, which corresponds to 31-32 T for $B_{c2}$ (0 K), the highest reported measured $B_{c2}$ for Nb$_3$Sn [54].

Table II. Comparison of Nb$_3$Sn fractions, $B_{c2}$, and pinning characteristics of the present best RRP, PIT and TT conductors, and the presumed limits for Nb$_3$Sn superconductors.

|  | Fine-grain Nb$_3$Sn fraction | $B_{c2}$ (4.2 K), T | Flux Pinning | |
| --- | --- | --- | --- | --- |
|  |  |  | Grain size | $F_p(B)$ peak |
| Present best RRP | ~60 % | 25-27 T | 120-150 nm | ~0.2$B_{c2}$ |
| Present best PIT, TT | 40-45 % | 25-27 T | 100-150 nm | ~0.2$B_{c2}$ |
| Limits or optimum | ~65 % | ~28 T | < 15 nm | 0.5$B_{c2}$ |

From Table II it is clear that the room for further improving Nb$_3$Sn fraction and $B_{c2}$ of state-of-the-art Nb$_3$Sn conductors (RRP type) is marginal. By optimizing chemistry, processing, and heat treatments of Nb$_3$Sn strands to push these two factors to their limits, it may be possible to further improve non-Cu $J_c$ of the state-of-the-art Nb$_3$Sn strands by up to 10-20%; however, further greater improvement along these lines seems not likely. It is also worth pointing out that an additional difficulty lies in that the above three factors are not independently influenced by the starting chemistry and processing of conductors. In many cases a modification to improve one of the factors may cause negative impacts to other factors. For example, increasing reaction temperature benefits $B_{c2}$ but undermines flux pinning; increasing Cu/Sn ratio in RRP, PIT and TT wires helps in reducing high-Sn intermediate phases and improving fine-grain Nb$_3$Sn fraction (Figure 9), but could be detrimental to stoichiometry and $B_{c2}$ (Section 5). Thus, there may be trade-offs between these factors, making it difficult to push all of them to their optimal levels simultaneously.

On the other hand, it is clear from Table II that there is still a lot of room for improving the pinning, since the optimal grain size is far below those of present strands. To achieve further significant improvement of $J_c$ (for instance, by 40-45% for RRP conductors as required by the FCC project as shown in Section 1), efforts should be made to improve flux pinning, particularly via introduction of APCs.

## 8. Summary

Improvement in $J_c$ of Nb$_3$Sn conductors is highly needed to reduce size and cost of magnets, particularly for circular proton accelerators. It is shown that due to large performance variation in series production of conductors, improvement of $J_c$ by 40-45% for present state-of-the-art RRP wires is required for FCC specification. With the record $J_c$ of Nb$_3$Sn strands plateaued for nearly two decades, new techniques are needed to lift $J_c$ out of stagnation. This article reviews the factors that determine the non-Cu $J_c$ of Nb$_3$Sn conductors, and discusses the prospects for further optimizing each factor. The non-Cu $J_c$ of Nb$_3$Sn strands is determined by three factors: current-carrying Nb$_3$Sn fraction in subelements, $B_{c2}$, and pinning capacity, with the latter two determining the layer $J_c$. An analytic model developed to explore the dependence of phase fractions on precursor ratios shows that the current-carrying Nb$_3$Sn fraction is at maximum as precursor ratios make the high-Sn intermediate phases amount drop to zero, and the limit of fine-grain Nb$_3$Sn fraction in Nb$_3$Sn subelements is estimated to be ~65%. Approaches to improving current-carrying Nb$_3$Sn fractions of PIT and TT wires are proposed. The diffusion reaction process for the formation of Nb$_3$Sn is analyzed in this article, which explores what controls stoichiometry of Nb$_3$Sn phase formed in the Sn source/Nb$_3$Sn/Nb system. This theory explains how Sn source influences Sn content gradient and growth rate of the Nb$_3$Sn layer, as well as why Ti tends to make Nb$_3$Sn layer stoichiometry more uniform. The theory could be used as a guide to improve Nb$_3$Sn stoichiometry and $B_{c2}$. The internal oxidation technique, which forms ZrO$_2$ nano inclusions in Nb$_3$Sn and significantly refines grain size and improves $J_c$, is feasible to be implemented in practical Nb$_3$Sn wires. Further work to optimize this technique is proposed. Finally, after summarizing all the factors, it is seen that the only opportunity for further significantly improving non-Cu $J_c$ of Nb$_3$Sn conductors lies in improving pinning, particularly by the introduction of APCs.

**Acknowledgements**

The author thanks Edward W. Collings for reviewing the language of this manuscript.